\def\paperTitle{HiScene: Creating Hierarchical 3D Scenes with Isometric View Generation}

\def\authorBlock{
    Wenqi Dong$^{1,2}$\footnotemark[1]\footnotemark[3] \qquad
    Bangbang Yang$^{2}$\footnotemark[1] \qquad
    Zesong Yang$^{1,2}$ \qquad
    Yuan Li$^{1}$ \\
    Tao Hu$^{2}$ \qquad
    Hujun Bao$^{1}$ \qquad
    Yuewen Ma$^{2}$ \qquad
    Zhaopeng Cui$^{1}$\footnotemark[2] \\
    \textsuperscript{1}Zhejiang University \qquad
    \textsuperscript{2}ByteDance \\
}

\newif\ifreview 
\newif\ifarxiv \newcommand{\arxiv}{\arxivtrue}
\newif\ifcamera 
\newif\ifrebuttal 

\arxiv

\pdfoutput=1
\documentclass[10pt,twocolumn,letterpaper]{article}
\ifreview \usepackage[review]{cvpr} \fi
\ifarxiv \usepackage[pagenumbers]{cvpr} \fi
\ifrebuttal \usepackage[rebuttal]{cvpr} \fi
\ifcamera \usepackage{cvpr} \fi

\usepackage{graphicx}	
\usepackage{amsmath}	
\usepackage{amssymb}	
\usepackage{booktabs}
\usepackage{times}
\usepackage{microtype}
\usepackage{epsfig}
\usepackage{caption}
\usepackage{float}
\usepackage{placeins}
\usepackage{color, colortbl}
\usepackage{stfloats}
\usepackage{enumitem}
\usepackage{tabularx}
\usepackage{xstring}
\usepackage{multirow}
\usepackage{xspace}
\usepackage{url}
\usepackage{subcaption}
\usepackage{xcolor}
\usepackage[hang,flushmargin]{footmisc}

\usepackage{amsfonts}
\usepackage{algorithm}
\usepackage{algorithmic}
\usepackage{amsmath}
\usepackage{xcolor}

\ifcamera \usepackage[accsupp]{axessibility} \fi

% ...

\ifarxiv  \fi

\newcommand{\R}[1]{{%
    \textbf{%
        \ifstrequal{#1}{1}{\textcolor{red}{R#1}}{%
        \ifstrequal{#1}{2}{\textcolor{blue}{R#1}}{%
        \ifstrequal{#1}{3}{\textcolor{magenta}{R#1}}{%
        \ifstrequal{#1}{4}{\textcolor{teal}{R#1}}{%
                           \textcolor{cyan}{R#1}%
        }}}}%
    }%
}}

\definecolor{url_color}{RGB}{42, 83, 163}

\usepackage{xr-hyper}

\makeatletter
\newcommand*{\addFileDependency}[1]{
  \typeout{(#1)}
  \@addtofilelist{#1}
  \IfFileExists{#1}{}{\typeout{No file #1.}}
}

\makeatother
\newcommand*{\myexternaldocument}[1]{
    \externaldocument{#1}
    \addFileDependency{#1.tex}
    \addFileDependency{#1.aux}
}

\definecolor{cvprblue}{rgb}{0.21,0.49,0.74}
\usepackage[pagebackref,breaklinks,colorlinks,allcolors=cvprblue]{hyperref}
\usepackage[capitalize]{cleveref}
\crefname{section}{Sec.}{Secs.}
\crefname{table}{Table}{Tables}
\crefname{figure}{Fig.}{Figs.}

\ifarxiv \crefname{appendix}{App.}{Apps.}
\else \crefname{appendix}{Suppl.}{Suppls.} \fi

\unless\ifarxiv \myexternaldocument{_supplementary} \fi

\begin{document}
\title{\paperTitle}
\author{\authorBlock}
\twocolumn[{
\renewcommand\twocolumn[1][]{#1}
\maketitle
    \includegraphics[width=\textwidth]{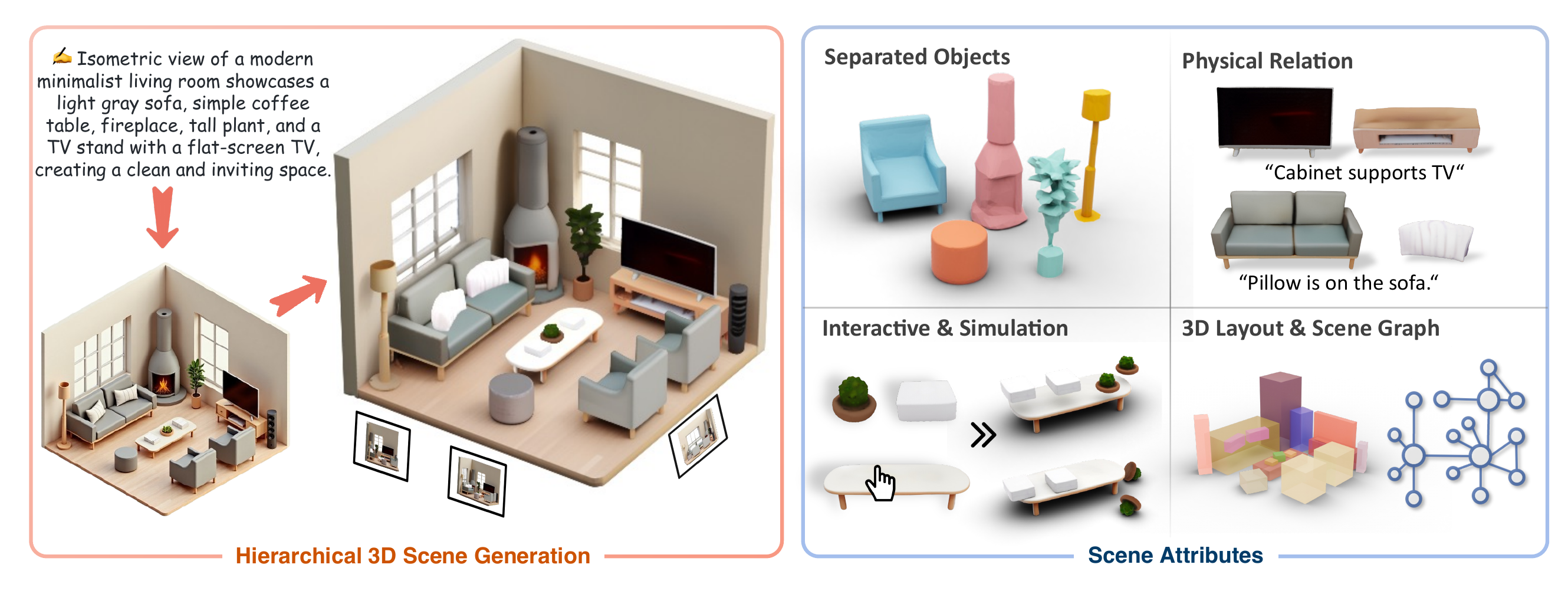}
    \vspace{-1.0em}
    \captionof{figure}{
    \textbf{HiScene} allows users to generate scene-level 3D assets with natural layout and appealing looking, while delivering compositional items for versatile applications such as interactive editing and simulation.
    }
    \label{fig:teaser}
}]

\renewcommand{\thefootnote}{\fnsymbol{footnote}}
\footnotetext[1]{Authors contributed equally.}
\footnotetext[2]{Corresponding author.}
\footnotetext[3]{Work done during an internship at PICO, ByteDance.}

\begin{abstract}
Scene-level 3D generation represents a critical frontier in multimedia and computer graphics, yet existing approaches either suffer from limited object categories or lack editing flexibility for interactive applications.
In this paper, we present HiScene, a novel hierarchical framework that bridges the gap between 2D image generation and 3D object generation and delivers high-fidelity scenes with compositional identities and aesthetic scene content.
Our key insight is treating scenes as hierarchical "objects" under isometric views, where a room functions as a complex object that can be further decomposed into manipulatable items.
This hierarchical approach enables us to generate 3D content that aligns with 2D representations while maintaining compositional structure.
To ensure completeness and spatial alignment of each decomposed instance, we develop a video-diffusion-based amodal completion technique that effectively handles occlusions and shadows between objects, and introduce shape prior injection to ensure spatial coherence within the scene.
Experimental results demonstrate that our method produces more natural object arrangements and complete object instances suitable for interactive applications, while maintaining physical plausibility and alignment with user inputs. More details at \urlstyle{tt}
\textcolor{url_color}{\url{https://zju3dv.github.io/hiscene/}}.
\end{abstract}

\vspace{-1.0em}
\section{Introduction}
\label{sec:intro}
Recently, we have witnessed remarkable breakthroughs of generative techniques in both 2D and 3D content creation, which empowers users to create stunning images and intricate 3D objects with simple text prompts.
However, extending such generation capabilities to scene-level 3D content—for instance, generating complete 3D room models with coherent geometry and appearance guided by user-provided text or images—remains a significant challenge.
Recent approaches typically rely on large language models (LLMs) with handcrafted rules to generate 3D scene layouts and place objects accordingly, yet these results often lack realism 
(e.g., exhibiting constrained object diversity and simplistic arrangements) 
due to LLMs' limited spatial understanding capabilities~\cite{wang2024architect}.
Others attempt to lift 2D images into 3D scenes using depth-based mesh deformation~\cite{hollein2023text2room, chung2023luciddreamer}. While these approaches offer category-free generation capabilities, they typically produce scenes as inseparable wholes, limiting interactive applications such as scene editing, object manipulation, and data curation for robotic semantic understanding.

Based on these observations, we believe an ideal scene generation method should have the following properties:
\textbf{1) Realistic layout \& assets}: it should generate scenes with natural object arrangements and diverse content beyond just a few simple objects from limited categories. The scenes should reflect real-world spatial relationships and object interactions.
\textbf{2) Compositional \& complete instances}: for interactive applications, each object in the scene should be a complete, intact 3D entity that can be individually manipulated, edited, or replaced without disrupting the entire scene.
\textbf{3) Spatial alignment \& plausibility}: the generated scene should faithfully represent the user's text or image prompt while maintaining physical coherence and plausibility in the 3D space (e.g., without distorted shapes, unrealistic proportions, or floating placement).

In this paper, we propose a novel hierarchical scene generation framework, named HiScene.
Rather than determining how scenes are built in 3D space with handcrafted rules, we leverage the complementary knowledge embedded in image generation models about how scenes should appear with aesthetic appeal and reasonable layout, and instantiate the concrete 3D representation that aligns with the image in a top-down manner.
Our key insight is that we can treat a scene as hierarchical level ``objects'' under the \textbf{isometric view}.
From the generator's perspective, a room can be seen as a complex object itself, while each individual item within the room can also be separately generated and manipulated.
By leveraging this hierarchical approach, HiScene bridges the gap between object-level and scene-level generation, producing complete scenes that benefit from pre-trained object-centric generation priors while maintaining compositional structure.
While the hierarchical scene generation approach is technically plausible, we identify several challenges in creating high-fidelity and compositional 3D scenes that we address in this paper.

\noindent\textbf{Hierarchical Scene Parsing.}
To bootstrap the hierarchical scene generation, we first initialize the entire scene with a pre-trained 3D generation model~\cite{xiang2024structuredtrellis} from the given isometric view image.
Once obtaining the complete 3D Gaussian representation~\cite{kerbl20233dgaussiansplatting} of the scene, 
a key challenge is to accurately isolate individual objects from the scene structure.
To address this, we implement a hierarchical scene parsing approach based on ``analysis by synthesis'' to identify distinct objects and prepare them for subsequent generative refinement.
Specifically, we render multi-view images and perform 3D identity segmentation using 2D segmentation priors enhanced with contrastive learning techniques~\cite{ying2024omniseg3d}.
Then, for each identified object, we render object-centric circular views and carefully identify occlusion regions, enabling us to understand the spatial relationships between objects and more effectively reconstruct complete object identities in the following steps.

\noindent\textbf{Instance Refinement with Video-diffusion-based Amodal Completion.}
A key challenge during identity refinement is that the rendered instance views often exhibit significant occlusions.
Despite advances in 3D object generation, reconstructing complete objects from occluded views remains ill-posed, while directly applying standard inpainting methods often produces implausible results due to limited object understanding (see Fig.~\ref{fig:comp_amodal}).
Moreover, the target instance might also include ambient shadow caused by the foreground occluder, which cannot be addressed with conventional inpainting frameworks.
To tackle this problem, we reformulate the instance refinement as a 2D amodal completion and 3D regeneration task, and propose a novel video-diffusion-based completion framework to handle it.
Specifically, our approach treats the amodal completion process as a temporal transition video effect, where occlusions gradually dissolve to reveal the complete object.
To enable this capability, we construct a specialized dataset for training video diffusion model to perform such completion transitions.
The temporal nature of our video-based completion effectively handles challenging cases including occlusion shadow removal, outperforming static image-based inpainting or completion methods by preserving structural coherence and producing more plausible results even in complex scenarios.

\noindent\textbf{Spatial Aligned Generation.}
Even with advanced 3D generation models to refine each segmented identity, ensuring spatial alignment between the refined object and its original placement is non-trivial due to the unposed nature of compressed latent~\cite{xiang2024structuredtrellis}. 
As a result, na\"ively applying refinement with amodally completed object views might produce objects with variant shapes, making them incompatible with the original scene layout.
To address this issue, we propose a shape prior injection mechanism that conditions the refinement stage of each identity.
Specifically, we first extract a geometric shape prior from a view-aligned generation method~\cite{xu2024instantmesh}, and use this aligned shape prior as the latent initialization for our refinement pipeline rather than starting from random noise.
This approach significantly reduces geometric ambiguity during refinement and ensures proper spatial alignment between the generated objects and the original scene context.

The contribution of the paper can be summarized as follows.
\textbf{1)}
We propose HiScene, a novel scene-level generation method that produces high-fidelity 3D scenes with compositional identities, natural scene arrangement and diverse content.
HiScene bridges the gap between 2D and 3D generation by leveraging isometric views, enabling effective hierarchical scene-level generation with pre-trained 3D object generation models.
\textbf{2)}
We develop an analysis-by-synthesis approach for scene parsing with zero-shot 3D semantic segmentation, and introduce a video-diffusion-based amodal completion method that effectively handles occlusions and ambient shadows during instance refinement.
\textbf{3)}
We design a spatial alignment mechanism with shape prior injection that ensures refined objects maintain proper geometric alignment with the original scene context, ensuring the object coherence and physical plausibility of the compositional 3D scene.
\textbf{4)}
Extensive experiments demonstrate the effectiveness of our approach, demonstrating the superior performance of video-diffusion-based amodal completion in handling complex occlusions and shadows, and the high-quality scene decomposition and object refinement across various challenging scenarios.

\section{Related Work}
\label{sec:related}

\subsection{3D Object Generation}
Motivated by the recent advances of diffusion techniques in 2D image generation, numerous works~\cite{poole2022dreamfusion, lin2023magic3d, wang2023scoresjc, metzer2023latent, chen2023fantasia3d, wang2023prolificdreamer, tang2023dreamgaussian, yi2024gaussiandreamer, liang2024luciddreamer} are being made to apply diffusion models to 3D object generation. DreamFusion~\cite{poole2022dreamfusion} first attempts to distill 2D gradient priors from the denoising process by employing score distillation sampling loss (SDS loss). Follow-up methods aim to enhance both the quality~\cite{lin2023magic3d, wang2023scoresjc, metzer2023latent, chen2023fantasia3d, wang2023prolificdreamer} and efficiency~\cite{tang2023dreamgaussian, yi2024gaussiandreamer,liang2024luciddreamer} of this method. 
Zero123~\cite{liu2023zero123} constructs paired viewpoint data on the large-scale dataset Objaverse~\cite{deitke2023objaverse} and fine-tunes the 2D latent diffusion model to achieve arbitrary novel view synthesis and 3D generation under single-image condition. Later works~\cite{tang2023emergentmvdiffusion, shi2023mvdream, liu2023syncdreamer, long2024wonder3d, shi2023zero123++} address the issue of cross view inconsistency in Zero123 by synchronously generating multiview images.
To enhance 3D object generation efficiency, Large Reconstruction Model (LRM)~\cite{hong2023lrm} employ a transformer-based architecture that directly reconstructs 3D objects from single-view image through feed-forward inference. Some methods~\cite{liu2024one2345pp, xu2024instantmesh, xu2024sparp, tang2024lgm, xu2024grm, li2023instant3d} incorporated multi-view information as input, significantly improving generation quality. Recent native 3D generation model~\cite{zhang20233dshape2vecset, zhang2024clay, xiang2024structuredtrellis, zhao2023michelangelo, li2024craftsman, chen20243dtopia, zhao2025hunyuan3d, lan2024gaussiananything} such as Shape2Vecset~\cite{zhang20233dshape2vecset}, CLAY~\cite{zhang2024clay}, and TRELLIS~\cite{xiang2024structuredtrellis} adopt a decoupled strategy, dividing the process into geometric structure generation and texture generation stages. These methods train generative model directly on 3D data rather than traditional multi-view data, substantially improving the geometric accuracy and consistency of generated results. Our method use existing native 3D generation model, treating scenes as hierarchical level of ``objects'' to ultimately achieve interactive scene generation.

\subsection{Text-to-3D Scene Generation}
Text-conditioned 3D scene generation has advanced rapidly in recent years.
Text2Room~\cite{hollein2023text2room} and follow-up works~\cite{chung2023luciddreamer, zhang2024text2nerf, yu2024wonderworld, yu2024wonderjourney, zhou2024dreamscene360} leverage latent diffusion models~\cite{rombach2022highldm} and monocular depth estimators~\cite{bae2022irondepth, bhat2023zoedepth} to iteratively generate textured 3D meshes or Gaussian Splatting scenes by inpainting from randomly sampled camera angles. However, these methods typically produce coupled scenes where object instances are difficult to isolate.
Some methods~\cite{paschalidou2021atiss, tang2024diffuscene, zhai2023commonscenes, zhai2024echoscene, lin2024instructscene, yang2021scene, ocal2024sceneteller} first generate 3D scene layouts, then obtain individual objects through retrieval or generation methods, placing them within the scene according to the generated layout. BlockFusion~\cite{wu2024blockfusion} and its follow-up methods~\cite{meng2024lt3sd, bokhovkin2024scenefactor} adopt a native 3D scene generation strategy, dividing the scene into multiple blocks and expanding these blocks into a complete scene during generation. However, these two types of methods are often limited to specific categories from training data.
Recent methods~\cite{li2024dreamscene, zhou2024gala3d, li2024discene, zhang2024towards} explore interactive general scene generation using Large Language Models (LLMs) to construct 3D layouts, while generating individual objects through Score Distillation Sampling-based optimization. However, LLMs still lack sufficient spatial understanding capabilities, making it difficult to generate complex and physically plausible 3D layout structures.
By contrast, our method adopts a top-down hierarchical generation that ensures global layout and appearance coherence while maintaining the separability of individual objects.

\subsection{Image Inpainting and Amodal Completion}
\label{sec:inpaint_amodal}
Image Inpainting and Amodal Completion are classic problems in computer vision. Image inpainting aims to restore masked regions in images with reasonable and natural content, requiring specification of the inpainting area. Traditional methods~\cite{bertalmio2000imageinpainting, criminisi2004regionfilling, peng2021generating, liu2021pd, zhao2021large} offen relied on auxiliary hand-engineered features with often poor results. Recent approaches~\cite{avrahami2023blended, avrahami2022blended, liu2023image, ju2024brushnet, jiang2025smarteraser, chen2024region} utilize diffusion models for text-guided completion, typically regenerating content in masked regions while preserving the rest of the image. In contrast, Amodal Completion aims to generate the complete form of an object from its visible view. Traditional methods~\cite{ke2021deep, ling2020variational, qi2019amodal, reddy2022walt, hsieh2023tracking, kar2015amodal, zhan2024amodal} focused on generating complete segmentation masks or predicting bounding boxes, while recent zero-shot diffusion-based method~\cite{ozguroglu2024pix2gestalt, xu2024amodal} further recover the complete content. We adopt Amodal Completion rather than Image Inpainting because Image Inpainting requires manually specified inpainting regions, whereas in Amodal Completion, the observed regions can be automatically obtained through existing segmentation networks, which is crucial for our automated generation of candidate images for each object during the hierarchical scene parsing stage.

\section{Method}
\label{sec:method}
\begin{figure*}[t!]
    \centering
    \includegraphics[width=1.0\linewidth, trim={0 5mm 0 0}, clip]{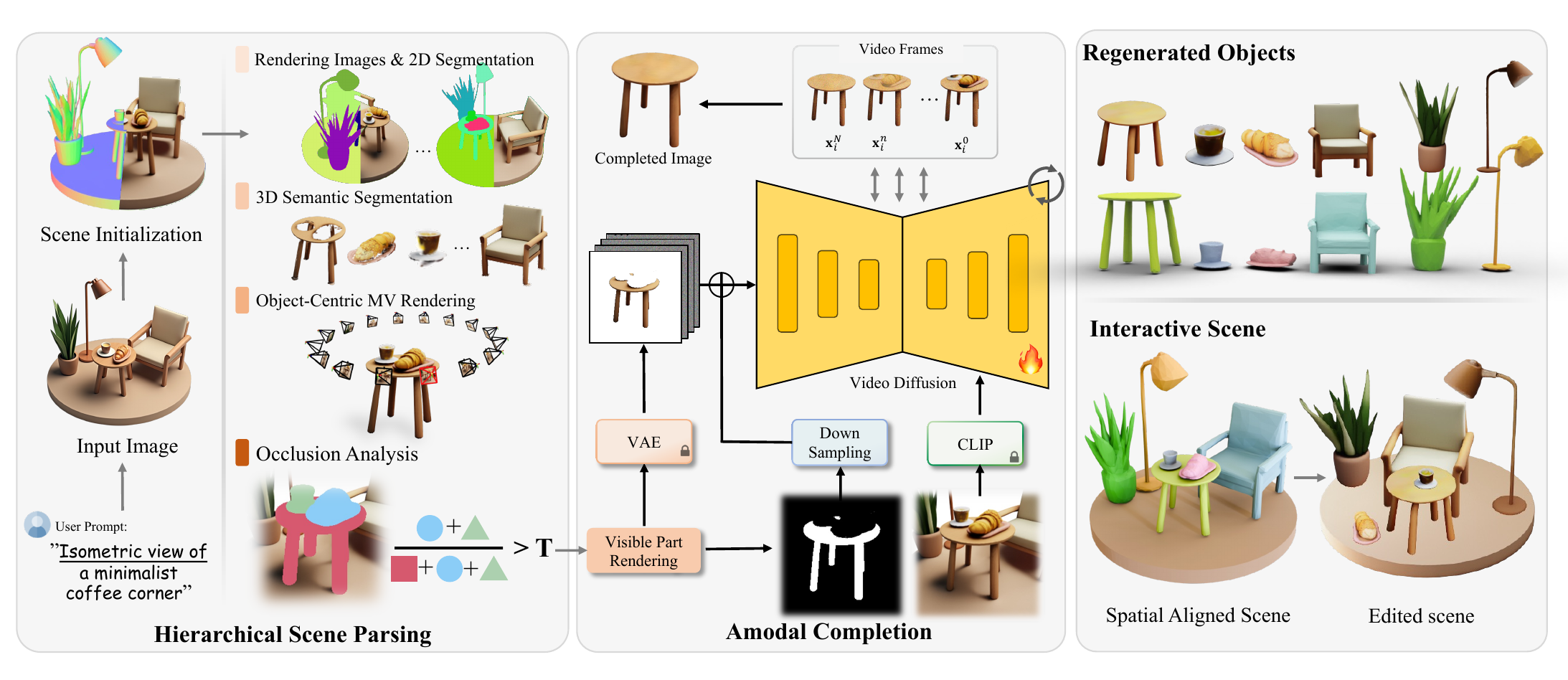}
    \caption{
    \textbf{Overview of HiScene.} Our hierarchical framework generates 3D scenes with compositional identities through three main stages. First, we create a 3D scene from a generated isometric view.
    Next, we perform scene parsing to obtain precise object segmentation, followed by multi-view rendering and detailed occlusion analysis for each identified instance.
    Finally, we apply our video-diffusion-based amodal completion to generate complete views of each instance, which serve as guidance for regenerating intact objects with proper spatial alignment in the scene.
    The resulting 3D scene features fully compositional identities, facilitating user-directed modifications like interactive scene editing.
    }
    \label{fig:pipeline}
\end{figure*}

We present HiScene, a novel hierarchical framework for generating compositional 3D scenes with intact and manipulatable objects.
As illustrated in Fig.~\ref{fig:pipeline}, we first initialize a 3D Gaussian Splatting scene from a generated isometric view. We then perform hierarchical scene parsing with semantic segmentation to identify distinct objects and obtain each object' multi-view rendering and occlusion analysis. Finally, we conduct video-diffusion-based amodal completion to address object occlusion and generate intact, spatially-aligned objects that enable interactive scene manipulation.

\subsection{Preliminary}
\label{method_preliminary}
\begin{figure}[t!]
    \centering
    \includegraphics[width=1.0\linewidth, trim={0 0 0 0}, clip]{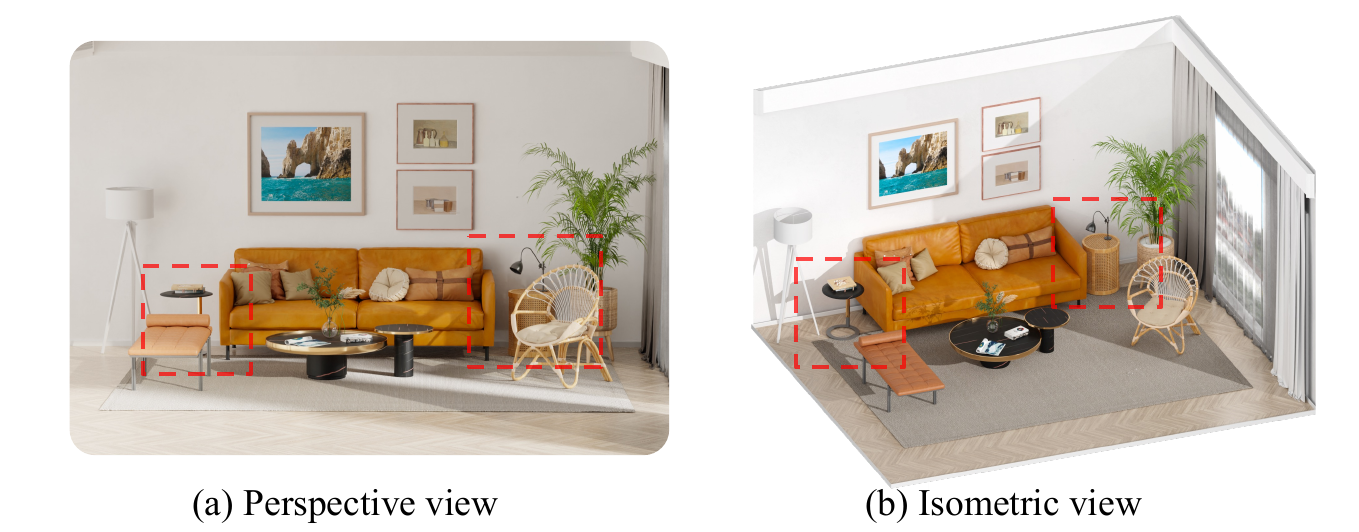}
    \caption
    {Comparison of perspective view and isometric view of a living room scene. Zoom in for more details.
    }
    \label{fig:method_isometric}
\end{figure}

\noindent\textbf{Isometric View.}
In computer graphics, isometric view is an orthographic projection method used to render 3D scenes into images.
Isometric view offers three key advantages for scene-level generation:
1) \textit{Distortion-free}: Unlike perspective projection, isometric view maintains consistent proportions without perspective distortion, ensuring accurate object representation.
2) \textit{Minimal-occlusion}: As shown in Figure~\ref{fig:method_isometric}, isometric view captures scenes from an elevated angle, revealing multiple faces of objects simultaneously while minimizing occlusion between scene elements.
3) \textit{Scene-as-object}: The unified representation of scenes in isometric view allows the entire scene to be treated as a cohesive entity, enabling direct generation with object-centric generative models.

\noindent\textbf{Native 3D Generation Model.}
We leverage native 3D generation models TRELLIS~\cite{xiang2024structuredtrellis} to bootstrap hierarchical scene generation.
TRELLIS introduces a unified Structured LATent (SLAT) representation $\boldsymbol{z}$ to characterize a 3D asset $\mathcal{O}$, as:
\begin{equation}
    \small \boldsymbol{z} = \{(\boldsymbol{z}_i,\boldsymbol{p}_i)\}_{i=1}^{L},\quad \boldsymbol{z}_i\in\mathbb{R}^C, \ \boldsymbol{p}_i\in \{0, 1,\ldots, N-1\} ^3, \label{eq:slate}
\end{equation}
TRELLIS employs a two-stage generation pipeline.
In the first stage, it generates the sparse structure $\{\boldsymbol{p}_i\}_{i=1}^{L}$ of $\mathcal{O}$ by first using a transformer model $\boldsymbol{\mathcal{G}}_{\mathrm{S}}$ to produce a low-resolution feature grid $\boldsymbol{S}\in\mathbb{R}^{D\times D\times D \times C_\mathrm{S}}$, followed by a latent feature decoder $\mathcal{D}_S$ to obtain a dense binary 3D grid $\boldsymbol{O} \in \{0,1\}^{N\times N\times N}$. The grid $\boldsymbol{O}$ is then converted to the set of 3D coordinates $\{\boldsymbol{p}_i\}_{i=1}^{L}$. In the second stage, TRELLIS uses another transformer model $\boldsymbol{\mathcal{G}}_{\mathrm{L}}$ to generate the corresponding structure features $\{\boldsymbol{z}_i\}_{i=1}^{L}$ for these coordinates. The complete SLAT representation $\boldsymbol{z}$ is then processed through specialized decoders ($\mathcal{D}_{\mathrm{NeRF}}$, $\mathcal{D}_{\mathrm{Mesh}}$, or $\mathcal{D}_{\mathrm{GS}}$) to produce the final 3D asset $\mathcal{O}$ in various formats (NeRF, meshes, or 3DGS).

\subsection{Hierarchical scene parsing}
\label{method_parsing}
We define an interactive scene $\mathcal{S}=\{\{\mathcal{O}_i, \mathcal{C}_i\}_{i=1}^N\}$ containing multiple separable complete objects $\{\mathcal{O}_i\}_{i=1}^N$ represented by 3DGS, with their configurations $\mathcal{C}_i = \{p_i, r_i, s_i, l_i\}$, where each configuration includes position $p_i \in \mathbb{R}^3$, rotation $r_i \in SO(3)$, scaling $s_i \in \mathbb{R}^3$, and semantic label $l_i$. 
To obtain this, as shown in Figure~\ref{fig:pipeline}, we first perform hierarchical scene parsing.

\noindent\textbf{Scene Initialization}
Given users' text prompt $T$, we first generate isometric view candidate images by prepending a fixed prefix \textit{"Isometric view of "} to the $T$.
Then, for the selected scene image $I_{scene}$, we obtain the initial scene representation $\mathcal{S}_0$ through TRELLIS. We adopt 3D Gaussian Splatting as our scene representation method. 

\noindent\textbf{3D Semantic Segmentation.}
We employ OmniSeg3D-GS~\cite{ying2024omniseg3d}, a contrastive learning-based semantic segmentation method for 3DGS. Specifically, after obtaining the initial representation $\mathcal{S}_0$, we render multi-view images $\mathcal{I} = \{I_j\}_{j=1}^{N_{VS}}$ from predefined viewpoints $\{V_j\}_{j=1}^{N_{VS}}$, where $N_{VS}$ is the number of scene views. Unlike OmniSeg3D-GS, which aims to achieve omniversal segmentation, our goal is to obtain separated objects $\{O_i\}_{i=1}^N$. Therefore, instead of using SAM~\cite{kirillov2023segment}, we employ EntitySeg~\cite{qi2022entityseg}, an off-the-shelf entity segmentation network, to generate class-agnostic instance-level 2D segmentation masks $\mathcal{M} = \{M_j\}_{j=1}^M$ for the multi-view images.

\noindent\textbf{Object-Centric Multiview Rendering.} 
As demonstrated in Figure~\ref{fig:pipeline}, due to the occlusion between objects in 3D scenes, after 3D segmentation, some objects (such as tables, croissants, coffee cups, etc.) typically appear incomplete. To further recover complete objects, 
we apply the following method to each object $\mathcal{O}_i$ in the scene: First, in the object's local coordinate system, we render multiview images $\{I_j^i\}_{j=1}^{N_{VO}}$ from predefined viewpoints $\{V_j^i\}_{j=1}^{N_{VO}}$, where $N_{VO}$ is the number of object views.
These rendered images will be served as candidate images for further object-centric instance regeneration.

\noindent\textbf{Occlusion Analysis.} 
\label{sec:occlusion_analysis}
After obtaining candidate images $\{I_j^i\}_{j=1}^{N_{VO}}$ for each object $\mathcal{O}_i$, we need to evaluate whether $\mathcal{O}_i$ is occluded by other objects in these images. For each candidate image $I_j^i$, we employ an advanced Vision Language Model (VLM) for occlusion detection. If occlusion is detected, we calculate the occlusion ratio of the target object. When this ratio exceeds threshold $\tau$, we apply Amodal Completion to recover the visual information in the occluded regions, thereby enhancing the quality and completeness of the candidate images.

\noindent For more details, please refer to the supplementary materials.
\begin{figure}[t!]
    \centering
    \includegraphics[width=1.0\linewidth, trim={0 0 0 0}, clip]{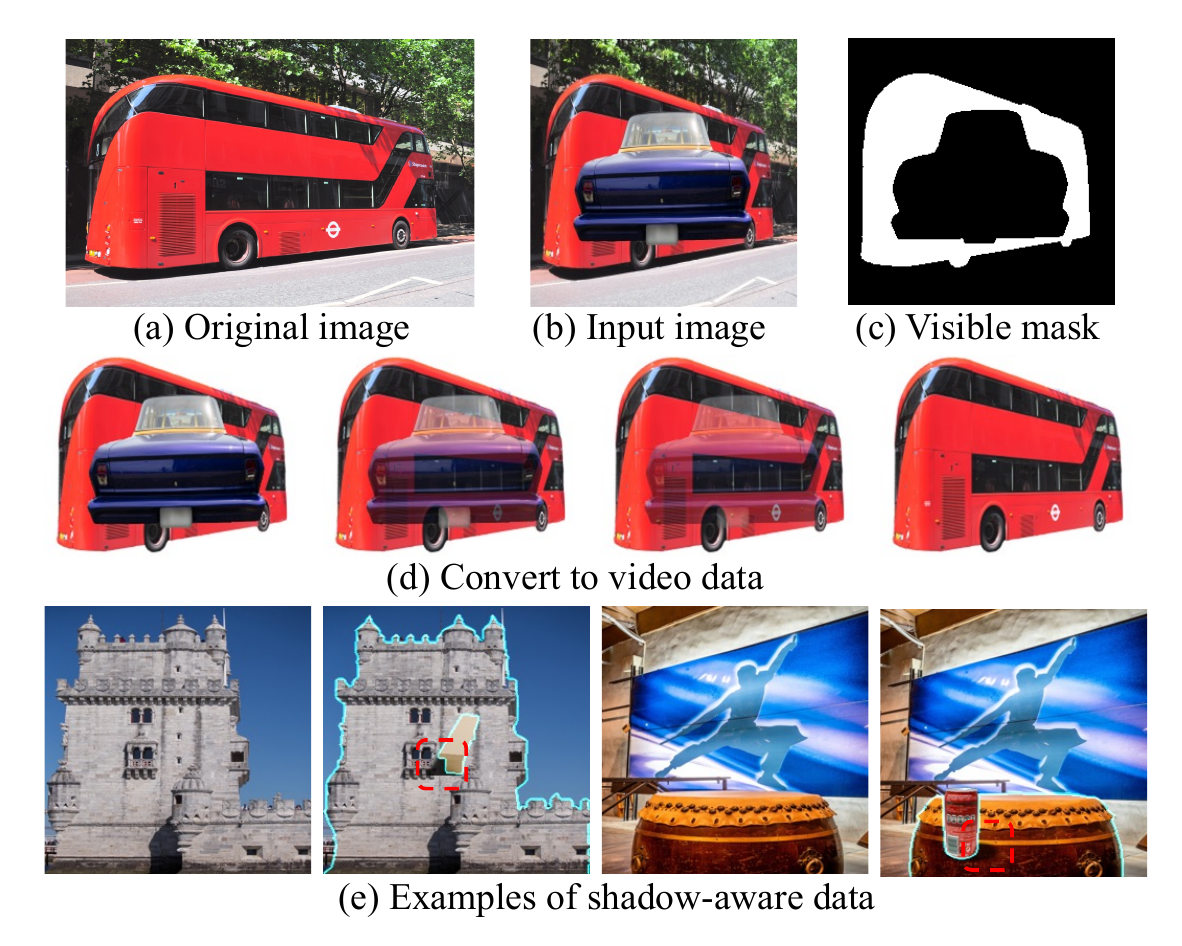}
    \caption
    {We present an data curation example of amodal completion, including original image (a), occluded input image (b), visible mask (c), and the linear blended video (d). We also present shadow-aware data examples (e).
    }
    \label{fig:method_amodal_dataset}
\end{figure}

\begin{figure*}[t!]
    \centering
    \includegraphics[width=1.0\linewidth, trim={0 0 0 0}, clip]{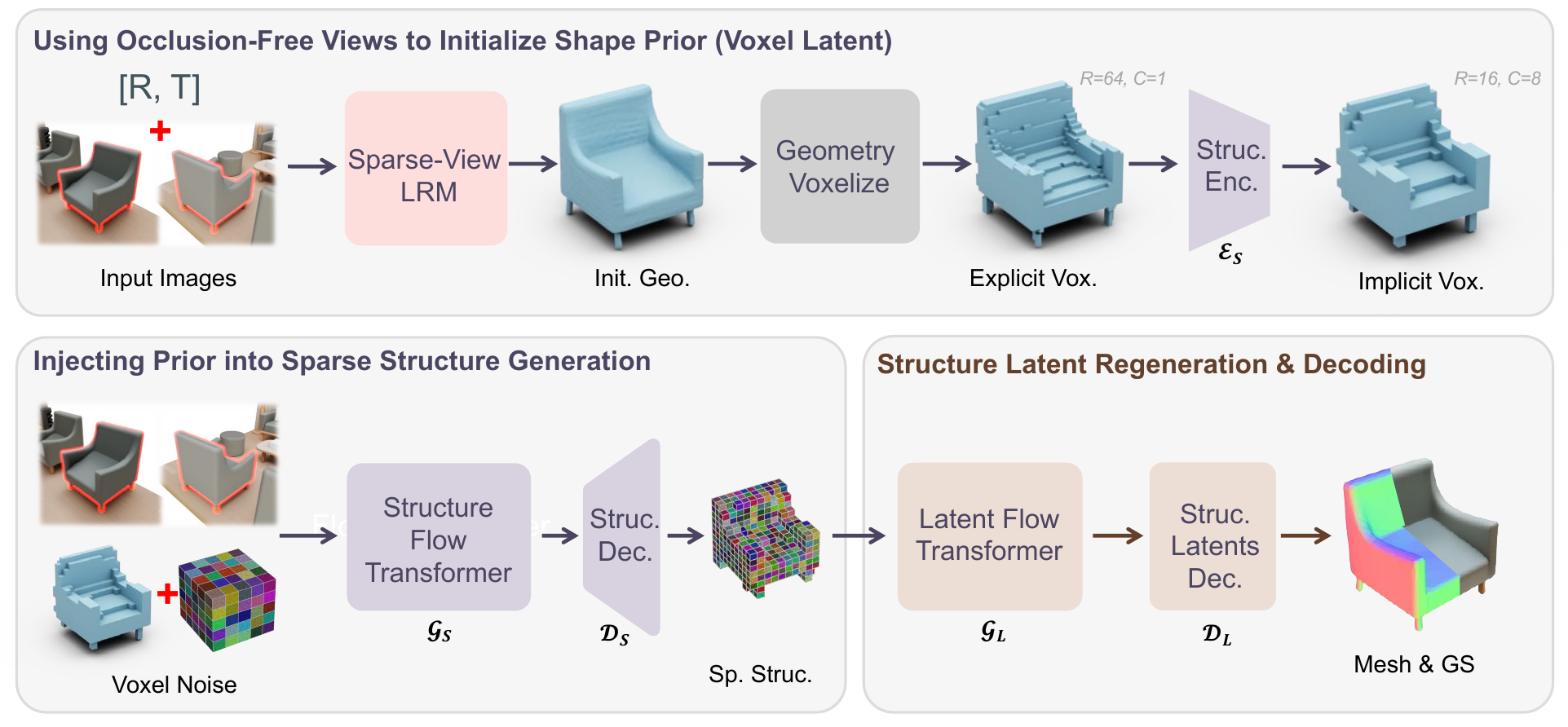}
    \caption{An illustration of Spatial Aligned Generation.
    We use sparse-view LRM to initialize spatial aligned shape prior (voxel latent), and inject this prior by initializing voxel noises upon it during native 3D generation, thus ensuring regenerated assets adhering the original scene.
    }
    \label{fig:geo_align}
\end{figure*}
\subsection{Amodal Completion}
\label{sec:method_amodal}
\noindent\textbf{Task Definition.} As discussed in the Section~\ref{sec:inpaint_amodal}, unlike image inpainting, amodal completion task aims to recover the complete and plausible form of the object when provided with an input image $I$ and visible mask $M$ (i.e., objects' visible part as shown in Figure~\ref{fig:method_amodal_dataset} (c)).
Unlike existing method Pix2gestalt\cite{ozguroglu2024pix2gestalt} that fine-tune image generation models, 
we define this task as a temporal transition video effect, where occlusions gradually dissolve to reveal the complete object.
Video models, trained on large-scale high-quality data, can learn temporal changes in the real world, thus possessing stronger prior knowledge that helps more accurately infer the complete form of occluded parts. 

\noindent\textbf{Dataset curation.} 
During object completion, apart from filling occluded parts, we need to remove notable visual artifacts caused by occlusion, such as shadows. Objects directly segmented from the SA-1B~\cite{kirillov2023sam} dataset cannot meet the requirements for constructing data with shadow effect. To this end, we constructed a large-scale dataset of objects with shadow occlusions using synthetic data. By filtering the Objaverse dataset~\cite{deitke2023objaverse}, we obtained 181K high-quality 3D objects. For each object, we utilized rigid body simulation to naturally place objects on the ground, ensuring realistic shadow effects. Additionally, we configured random lighting setups and employed the path-tracing renderer to generate 468K synthetic images. Integrating the data from Pix2gestalt, we ultimately built a training set comprising 1.32 million image pairs.
Then, we convert image pair into the required video data. As shown in Figure~\ref{fig:method_amodal_dataset} (d), we adopt a linear blending approach to put foreground object over the complete objects in image planes.

As shown in Figure 2, we employ the Image-to-Video model Stable Video Diffusion \cite{blattmann2023stablevideo}. Specifically, the visible part $I_{vis}$ is encoded through a VAE and used as the first frame input, concatenated with the downsampled visible region mask $M$ along the feature dimension. The whole image $I$ is processed through CLIP to extract features, which are then injected into the model via cross-attention. During training, we add noise $\epsilon\sim\mathcal{N}(0,I)$ to the original data $x_0$ to obtain $x_t = \sqrt{\alpha_t}x_{t-1} + \sqrt{1-\alpha_t}\epsilon$, and then use the network to predict the noise $\epsilon_\theta(x_t,t)$, minimizing the following loss function:
\begin{equation}
    \mathcal{L}_{dm} = \mathbb{E}_{x, \epsilon \sim \mathcal{N}(0,1), t}\left[\|\epsilon - \epsilon_\theta(x_t, t, I, I_{vis}, M)\|^2\right],
\end{equation}

Once the Amodal completion model is trained, we apply it to object candidate images with occlusion, which recovers intact image inputs for Spatial Aligned Generation stage.

\subsection{Spatial Aligned Generation}
\label{sec:spatial_align}
After obtaining objects' occlusion-free views, we aim to regenerate each object to achieve intact instances while preserving their original scale and poses. However, directly applying regeneration using native 3D generative models~\cite{xiang2024structuredtrellis} $F_{\text{Native}}$ often results in canonical objects that lose alignment with the original scene context.
To address this limitation, we propose injecting spatially aligned shape priors derived from multi-view large reconstruction models $F_{\text{LRM}}$~\cite{xu2024instantmesh} (refer to it as LRM for clarity) into the native 3D generation process.

As shown in Figure~\ref{fig:geo_align}, for each incomplete objects $\mathcal{O}_i$, we first use $F_{\text{LRM}}$ to reconstruct an initial geometric structure $S_{\text{init}}$ from the input observation views and their corresponding camera parameters $[r_i, p_i]$, and obtain an explicit voxel representation $V$ through voxelization. Subsequently, we employ the encoder $\mathcal{E}_S$ from the TRELLIS structure generation stage to compress $V$ into a low-dimensional latent feature $S_{\text{implicit}}$. This coarse 3D structure representation serves as guidance for the subsequent refinement process. Specifically, we use $S_{\text{implicit}}$ as the voxel latent initialization and add noise corresponding to an intermediate timestep $t$ of the rectified flow model (typically choosing $t \in [0.2, 0.4]$), rather than starting generation from pure random noise. The generator $\boldsymbol{\mathcal{G}}_{\mathrm{S}}$ then produces an optimized structure $\widehat{S}$. Under the guidance of $S_{\text{implicit}}$, the finally generated $\widehat{S}$ not only preserves the configuration of the initial geometry $M$ but also significantly enhances geometric details and texture quality.

\section{Experiments}
\label{sec:experiment}
In this section, we first demonstrate method's capabilities of text to interactive 3D scene generation in Sec.~\ref{ssec:exp_text_to_3d}. Next, we evaluate the effectiveness of our method in amodal completion task in Sec.~\ref{ssec:exp_amodal}. Finally, we conduct ablation studies to analyze the different components within our framework in Sec.~\ref{ssec:exp_ablation}.

\subsection{Interactive Scene Generation}
\label{ssec:exp_text_to_3d}
\begin{figure*}[t!]
    \centering
    \includegraphics[width=1.0\linewidth, trim={0 0 0 0}, clip]{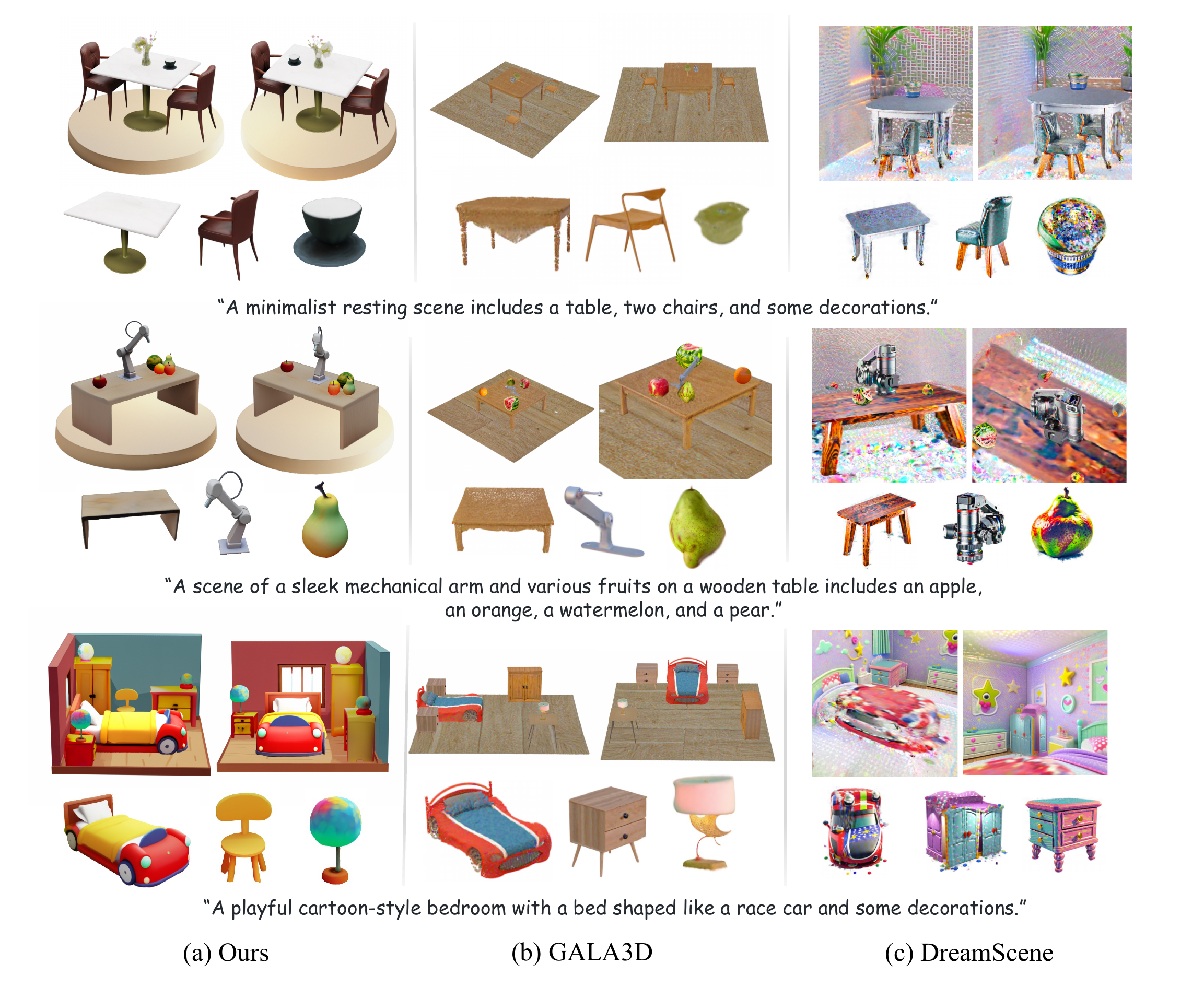}
    \caption{
    We compare the Interactive Scene 3D generation with GALA3D and DreamScene.
    }
    \label{fig:exp_scene}

\end{figure*}

We compare our method with two state-of-the-art decoupable scene generation methods: GALA3D~\cite{zhou2024gala3d} and DreamScene~\cite{li2024dreamscene}.
GALA3D employs large language models for generating initial layouts, integrates layout-guided Gaussian representation and adaptive geometry control, and utilizes a compositional optimization mechanism. DreamScene introduces Formation Pattern Sampling (FPS) to balance semantic information and shape consistency, and a three-stage camera sampling strategy to improve the quality of scene generation.
However, both methods require predefined 3D layouts as input, which presents a significant barrier for novice users who may find creating reasonable layouts challenging. Large language models often make errors in layout generation as well. As shown in Figure~\ref{fig:exp_scene}, our method addresses these limitations by providing a more intuitive and user-friendly approach to 3D scene generation without requiring explicit layout specifications.

\noindent{\textbf{Qualitative Comparison}} In Figure~\ref{fig:exp_scene}, we present both complete generated scenes and individual objects. As shown, scenes and objects generated by GALA3D and DreamScene exhibit artifacts. The layouts produced by these methods often violate physical constraints and common sense spatial relationships. Additionally, individual objects frequently suffer from oversaturation and the Janus problem. In contrast, our method generates complex yet plausible scenes with individual objects of significantly higher quality compared to the other approaches.

\noindent{\textbf{Quantitative Analysis}} To quantitatively evaluate our method, we employ CLIP Score to assess text-scene alignment, and use ImageReward and Aesthetic Score to evaluate the overall generation quality. As shown in Table~\ref{tab:comp_scene_user_study}, our method achieves the best overall performance. DreamScene's more severe multi-face object issues negatively impact its overall scores. Our approach demonstrates superior performance across all metrics, confirming the effectiveness of our layout-free scene generation paradigm.

\noindent{\textbf{User Study}} We also conducted a user study to compare our method against existing approaches. The evaluation focused on two aspects: text-scene alignment and overall quality. We collected 12 different scenes and asked 20 users to rate them on a scale from 1 to 3, with higher scores indicating better results. As shown in Table\ref{tab:comp_scene_user_study}, our method achieved the highest ratings, confirming the superior performance of our approach from a human perception perspective.

\begin{table}[
!t
]
    \setlength{\tabcolsep}{1.4pt}
    \caption{We perform quantitative evaluation and user studies
on the Interactive Scene Generation task.} 
    \centering
    \begin{tabular}{clccc}
        \toprule
         Method & Ours & GALA3D~\cite{zhou2024gala3d} & DreamScene~\cite{li2024dreamscene}  \\ 
        
        \midrule
        $\uparrow$ Aesthe. Score~\cite{murray2012ava} & \textbf{5.46} & 4.74 & 4.71 \\
        $\uparrow$ ImageReward~\cite{wu2023qalign} & ~\textbf{-0.28} & -1.67 & -0.73 \\
        $\uparrow$ CLIP Score$_\%$~\cite{radford2021learningclip} & ~\textbf{26.07} & 23.50 & 21.91 \\
        \midrule
        $\uparrow$ Matching Degree & \textbf{2.90} & 1.76 & 1.40 \\
        $\uparrow$ Overall Quality & \textbf{2.76} & 1.75 & 1.73 \\
        \bottomrule
    \end{tabular}

    \label{tab:comp_scene_user_study}
\end{table}

\subsection{Amodal Completion}
\label{ssec:exp_amodal}

\begin{figure*}[t!]
    \vspace{-1em}
    \centering
    \includegraphics[width=1.0\linewidth, trim={0 0 0 0}, clip]{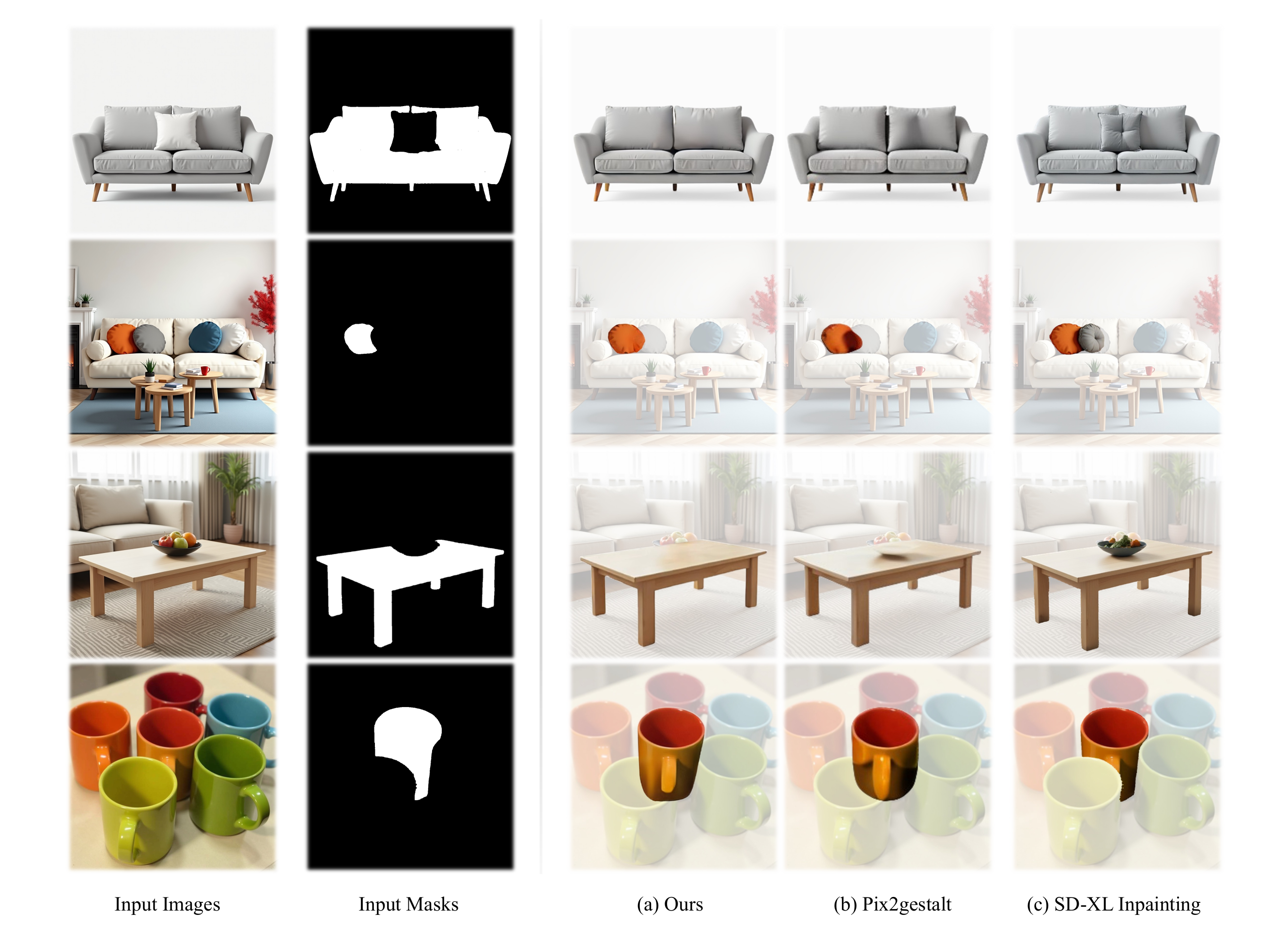}
    \vspace{-2em}
    \caption{
    \textbf{In-the-wild Amodal Completion and Segmentation.}
    }
    \label{fig:comp_amodal}
\end{figure*}
We assess the performance of amodal segmentation with existing zero-shot methods. Following~\cite{zhan2020self, ozguroglu2024pix2gestalt}, we evaluate segmentations on Amodal COCO (COCO-A)~\cite{zhu2017semanticcocoa} and Amodal Berkeley Segmentation (BSDS-A)~\cite{martin2001database} datasets using mean intersection-over-union (mIoU).  The COCO-A dataset offers 13,000 amodal annotations across 2,500 images, while the BSDS-A dataset includes 650 objects from 200 images. For both datasets, we evaluate methods that take an image and a (modal) mask of the visible portion of an object as input, and produce an amodal mask representing the full extent of the object. We use the same method as in pix2gestalt~\cite{ozguroglu2024pix2gestalt} to convert the amodal completions into semantic masks.

We compared our approach with the state-of-the-art method pix2gestalt and two other zero-shot methods, as shown in Table~\ref{tab:amodalseg}. Our method achieves state-of-the-art performance on both datasets, demonstrating that our video model-based approach more effectively recovers occluded objects, resulting in superior segmentation outcomes.
We also conducted qualitative experiments on everyday scenes, as illustrated in Figure~\ref{fig:comp_amodal}. Our method successfully recovers occluded objects and effectively removes shadows caused by occlusions. While pix2gestalt can reconstruct reasonable shapes, it often produces darkened textures in shadowed regions, likely due to the absence of shadow considerations in its training data. SD-XL Inpainting tends to be influenced by mask boundaries, frequently generating completions that conform to the mask but are semantically unreasonable.

\begin{figure}[h!]
    \centering
    \includegraphics[width=1.0\linewidth, trim={0 0 0 0}, clip]{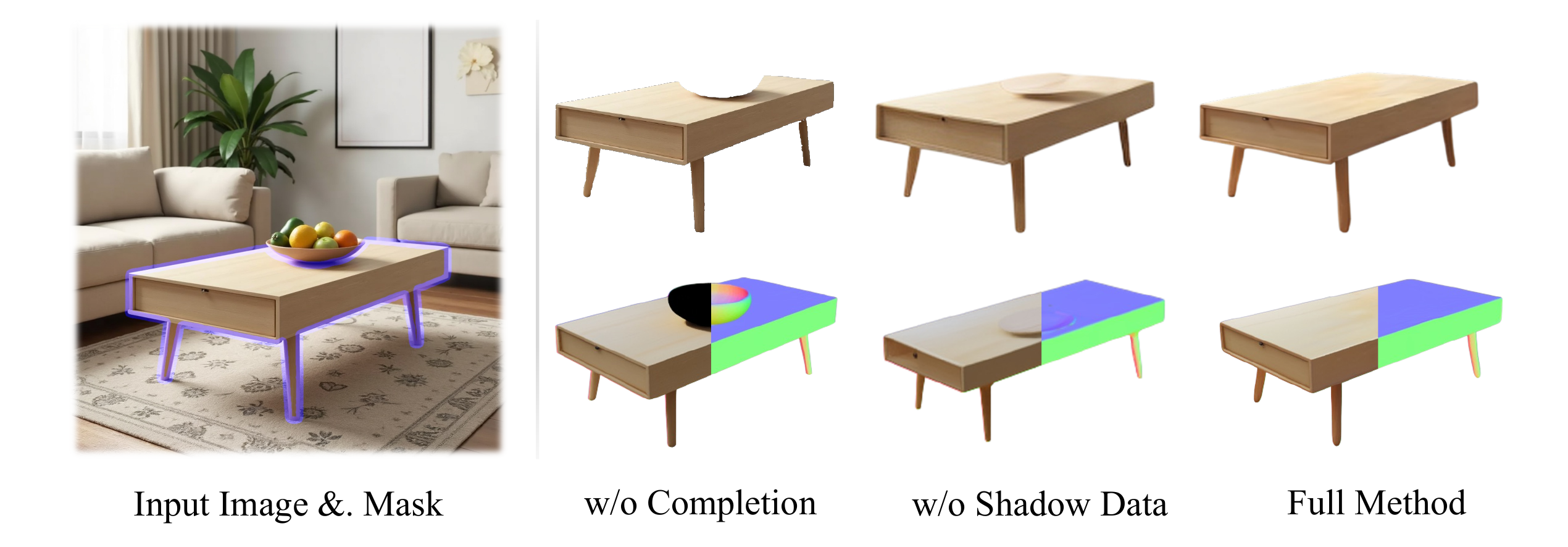}
    \caption
    {We analyzed the necessity of shadow-aware amodal completion.
    }
    \label{fig:ablation_recon}
    \vspace{-1.0em}
\end{figure}

\begin{table}[!t]
    \setlength{\tabcolsep}{12pt}
    \caption{Comparisons with zero-shot methods.}
    \centering 
    \begin{tabular}{clcc}
        \toprule
        Zero-shot Method & COCO-A & BSDS-A \\ 
        \midrule
        
         SAM~\cite{kirillov2023segment} & 60.27 & 60.20 \\
         SD-XL Inpainting~\cite{podell2023sdxl} & 70.08 & 66.57 \\
        
        Pix2gestalt~\cite{ozguroglu2024pix2gestalt} & 82.59 & 79.59 \\
        Ours & \textbf{83.84} & \textbf{79.80} \\
        \bottomrule
    \end{tabular}
    \label{tab:amodalseg}
\end{table}

\subsection{Ablation Studies}
\label{ssec:exp_ablation}
\textbf{Image vs. video model in amodal completion.} 
To qualitatively compare the capabilities of image and video models, we trained both types of models using data constructed by Pix2gestalt on the SA-1B dataset~\cite{kirillov2023segment}. We evaluated the overall quality of the generated completions using Aesthetic Score, Q-Align IAA, and IQA metrics, and measured text-image alignment using CLIP Score. As shown in Table~\ref{tab:ablation_amodal}, under the same data settings, the video model outperforms the image model across all metrics. We attribute this superior performance to the video model's powerful prior knowledge of object continuity and temporal consistency, which enables it to better understand and complete occluded objects with more coherent and realistic results.

\noindent\textbf{Shadow-aware completion for object generation.}
As illustrated in Figure~\ref{fig:ablation_recon}, our observation of the chair is incomplete due to occlusions. To demonstrate the importance of proper amodal completion, we first conducted an ablation experiment where we attempted object generation without amodal completion. Since object generation models are trained on complete observations, when presented with partial inputs, the model generates incorrect geometric structures based on the incomplete mask contours, often resulting in black textures in the missing regions. Similarly, when amodal completion is performed but shadow artifacts remain, the generated results still exhibit the aforementioned black geometric errors. Our shadow-aware amodal completion method effectively addresses these issues by properly handling both occlusions and shadows, resulting in geometrically accurate and visually coherent object reconstructions.

\begin{table}[h]
    \setlength{\tabcolsep}{5pt}
    \caption{We evaluated the effectiveness of the video model.} 
    \centering
    \begin{tabular}{clcclcc}
        \toprule
        \multirow{2}{*}{Datasets \& Method} & \multicolumn{2}{c}{COCO-A} & \multicolumn{2}{c}{BSDS-A} \\ 
        \cmidrule(r){2-3} \cmidrule(r){4-5}
        & I2I & I2V & I2I & I2V\\ 
        \midrule
        $\uparrow$ Aesthe. Score~\cite{murray2012ava} & 4.16 & \textbf{4.30} & 4.17 & \textbf{4.38} \\
        $\uparrow$ Q-Align IAA$_\%$~\cite{wu2023qalign} & 12.65 & \textbf{23.48} & 14.59 & \textbf{23.46} \\
        $\uparrow$ Q-Align IQA$_\%$~\cite{wu2023qalign} & 35.43 & \textbf{45.80} & 35.06 & \textbf{44.42} \\
        $\uparrow$ CLIP Score$_\%$~\cite{radford2021learningclip} & 20.27 & \textbf{20.78} & 20.63 & \textbf{21.20} \\
        \bottomrule
    \end{tabular}
    \label{tab:ablation_amodal}
\end{table}

\begin{figure}[h!]
    \centering
    \includegraphics[width=1.0\linewidth, trim={0 0 0 0}, clip]{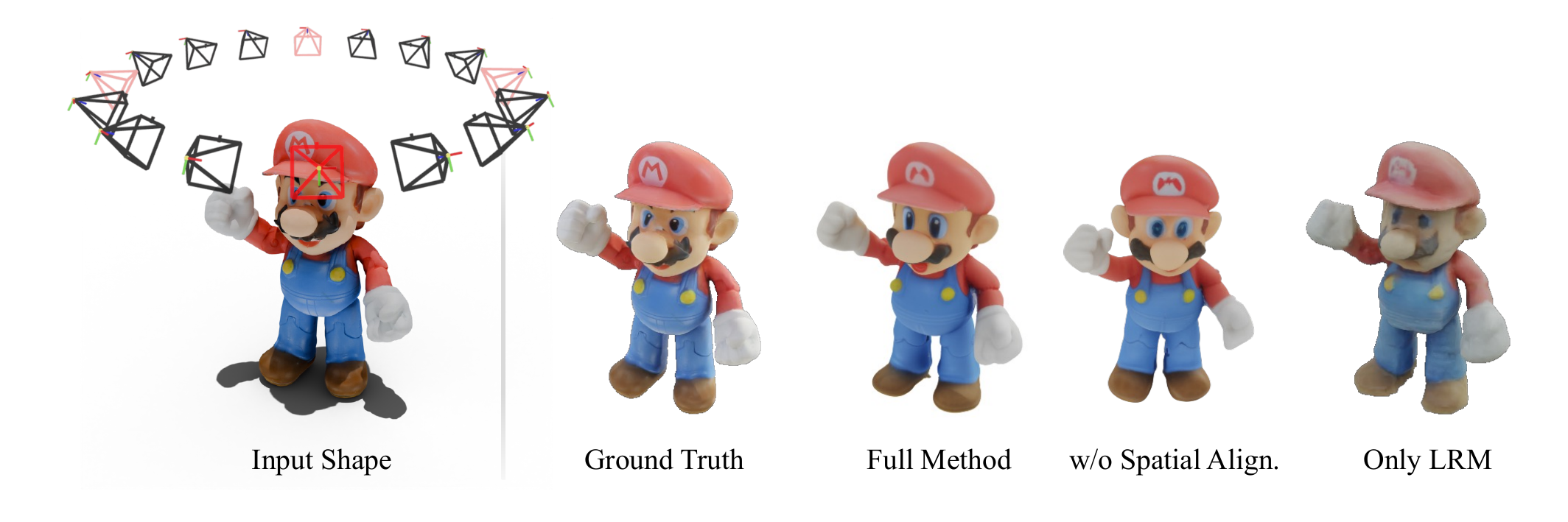}
    \caption
    {We analyze the effectiveness of Spatial Aligned Generation. 
    }
    \label{fig:ablation_spatial_aligned}
\end{figure}

\noindent\textbf{Spatial Alignment.}
We finally evaluate the effectiveness of our spatial aligned generation by comparing it with two alternatives: direct native 3D generation ($F_{\text{Native}}$ only) and standalone LRM generation ($F_{\text{LRM}}$ only).
As shown in Figure~\ref{fig:ablation_spatial_aligned}, without spatial alignment, native 3D generation produces objects with incorrect orientation and positioning relative to the ground-truth instance, while LRM generation alone results in compromised appearance fidelity.
By leveraging LRM's spatial alignment capabilities as a shape prior for native 3D generation, our approach achieves precise scale and pose matching with the original scene while preserving rich appearance details and visual quality.

\section{Conclusion}
\label{sec:conclusion}
In this paper, we have presented HiScene, a novel hierarchical framework for generating compositional 3D scenes.
By treating scenes as hierarchical compositions of objects under isometric views, we enable effective scene-level synthesis using pretrained object generation models.
To ensure completeness of each object identities, we use video-diffusion-based amodal completion and spatial alignment to aid the regeneration, ensuring spatial coherence within the scene.

\noindent
\textbf{Limitations and future works.}
Despite our advances, scenes generated by HiScene have textures with baked lighting, lacking PBR materials for modern rendering pipelines. Future work will focus on training the generative model to generate scenes that support PBR textures.

\clearpage

{\small
\bibliographystyle{ieeenat_fullname}
\bibliography{references}
}

\ifarxiv \clearpage  \appendix \maketitlesupplementary
In this supplementary material, we describe more training details of Hierarchical Scene Parsing in Sec.~\ref{sec:hierarchical_parsing} and Amodal Completion in Sec.~\ref{sec:supp_amodal}. We provide more details on Spatial Aligned Generation in Sec.~\ref{supp_spatial_align} and Sec.~\ref{supp_background}. We provide detailed explanation of our user study in Sec.~\ref{sup:userstudy}. We also conducted Runtime Evaluation in Sec.~\ref{sec:runtime_evaluation}. More qualitative results can be found in our supplementary video. Source code and dataset will be released upon the acceptance of this paper.

\section{Hierarchical Scene Parsing Details}
\label{sec:hierarchical_parsing}
\subsection{Predefined Viewpoints}
\begin{figure}[h!]
    \centering
    \includegraphics[width=1.0\linewidth, trim={0 0 0 0}, clip]{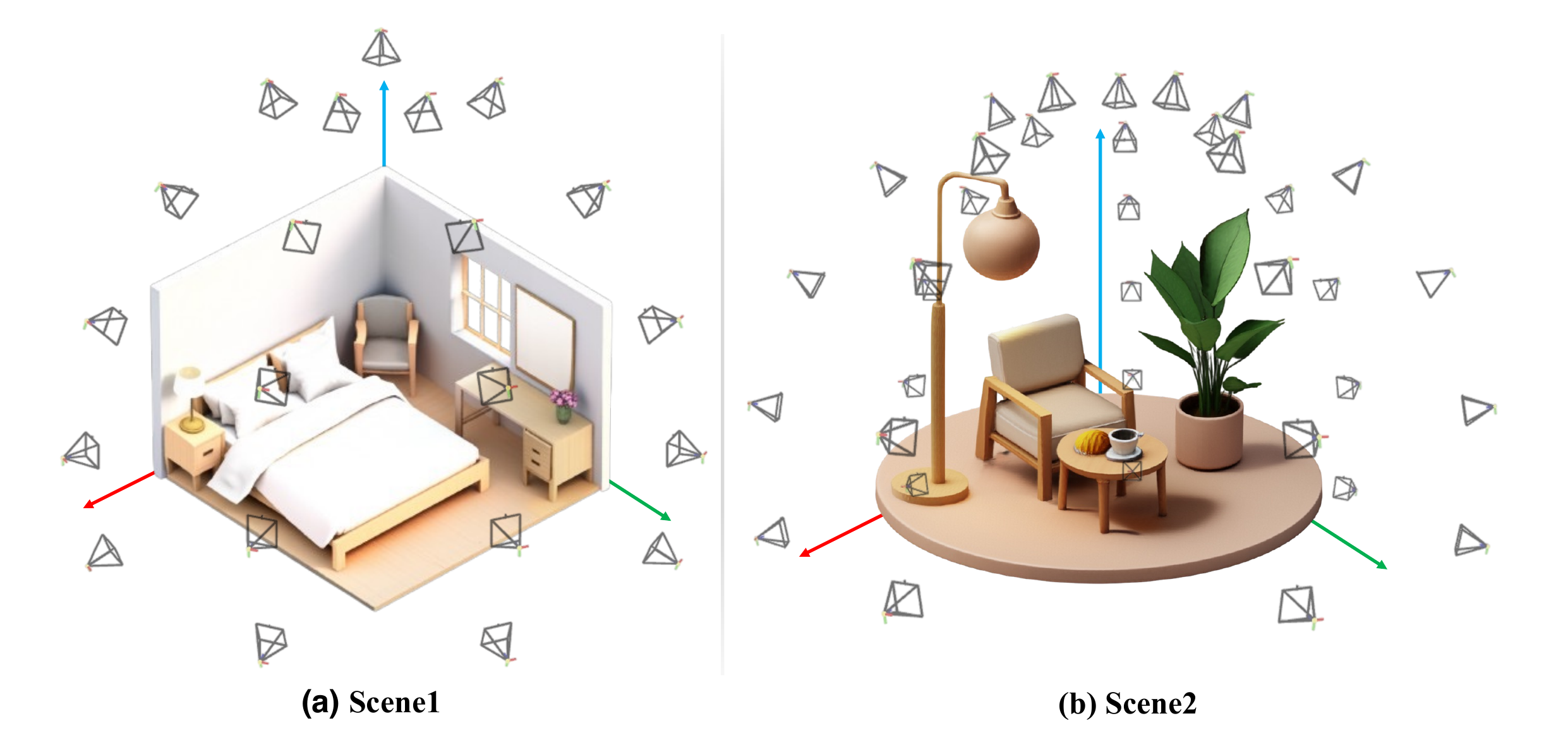}
    \caption{
        We provide examples of predefined viewpoints.
    }
    \label{fig:campose}
\end{figure}
In the 3D Semantic Segmentation stage, we render multi-view images of the initial scene $S_0$ from predefined viewpoints $V$. As shown in the Figure~\ref{fig:campose}, for scenes of type (a), we position camera viewpoints $V$ on a sphere with a fixed radius, with azimuth angles uniformly sampled at 8 points within $[-5^{\circ}, 95^{\circ}]$ and elevation angles uniformly sampled at 5 points within $[-10^{\circ}, 90^{\circ}]$. For scenes of type (b), the azimuth angles span $[0^{\circ}, 360^{\circ})$. The camera viewpoint set can be represented as $V = \{(r, \theta_i, \phi_j) \mid \theta_i \in \Theta, \phi_j \in \Phi\}$, where $r=2$ is the sphere radius, and $\Theta$ and $\Phi$ are the sampling sets for azimuth and elevation angles, respectively. Through experimental validation, we found that the specific configuration and number of rendering viewpoints do not significantly affect 3D segmentation performance.

As shown in Figure 2 of the main paper, during the object-centric novel view synthesis stage, we fix the elevation angle at $\phi=30^{\circ}$ and uniformly select 16 viewpoints with azimuth angles $\theta \in [0^{\circ}, 360^{\circ})$. All cameras are directed toward the center of the target object.

\subsection{3D Semantic Segmentation}
Existing methods~\cite{ye2024gaussiangrouping,choi2024clickgs, ying2024omniseg3d, koch2024open3dsg} support 3DGS semantic segmentation, and we adopt OmniSeg3D-GS~\cite{ying2024omniseg3d}. Given a collection of multi-view images $\{\mathbf{I}_i\}_{i=1}^M$ with their corresponding 2D semantic masks $\{\mathbf{M}_i\}_{i=1}^M$. OmniSeg3D-GS lifts class-agnostic 2D segmentation to 3D space through contrastive learning. To obtain complete 3D instance segmentation results, we employ EntitySeg~\cite{qi2022entityseg} for 2D segmentation. 

First, each Gaussian primitive in the initialized scene $\mathcal{S}_0$ is assigned an optimizable feature vector $\mathbf{h}_g \in \mathbb{R}^{16}$. The optimization phase follows an iterative approach, randomly selecting an image in each iteration and sampling $N$ points from it, while determining the patch id for each point. Subsequently, differentiable rendering techniques are used to compute the render features $\{\mathbf{f_i}\}(i\in[1,N])$ for each sampled point. Within the contrastive learning framework, points sharing the same patch id are treated as positive samples, while the remaining points are considered negative samples.

To enhance computational efficiency and ensure stable convergence, OmniSeg3D-GS applies a contrastive clustering method~\cite{li2020prototypical}. For a set of point features sharing the same patch id $i$, defined as cluster $\{\mathbf{f}^i\}$, with mean feature representation $\mathbf{\bar{f}}^{i}$. The contrastive clustering loss is defined as:
\begin{equation}
    \mathcal{L}_{CC}=-\frac{1}{N_p}\sum_{i=1}^{N_p}\sum_{j=1}^{|\{\mathbf{f}^i\}|}\log{\frac{\exp(\mathbf{f}^{i}_{j}\cdot \bar{\mathbf{f}}^i/ \phi_i)}{\sum_{k=1}^{N_p}\exp(\mathbf{f}^{i}_{j}\cdot \bar{\mathbf{f}}^k/ \phi_k)}},
    \label{eq:cc}
\end{equation}
where $\mathbf{f}_j^i$ represents the render feature with point index $j$ and patch id $i$. $N_p$ represents the total number of patch ids, and $\phi_i$ is the temperature parameter for the cluster, used to balance cluster size and variance, calculated as:
\begin{equation}
    \phi_i=\frac{\sum_{j=1}^{n_i}||\mathbf{f}^i_j-\bar{\mathbf{f}}^i||_2}{n_i\log(n_i+\alpha)}, \quad n_i=|\{\mathbf{f}^i\}|
\end{equation}
where $\alpha=10$ is a smoothing parameter that prevents small clusters from producing excessively large $\phi_i$ values. For more details, please refer to the original paper.

\subsection{Occlusion Analysis}
To achieve better generation results, the image condition for LRM and TREELIS models should ideally include complete multi-view observations of the target object. In the Predefined Viewpoints setting, we render 16 candidate images around the object with a 360-degree view. We uniformly divide the 360-degree view into 4 regions $\mathcal{R} = \{R_1, R_2, R_3, R_4\}$, with each region $R_i$ containing 4 images.
For each region $R_i$, we randomly select one image and use a vision-language model (VLM) ~\cite{achiam2023gpt4v} to determine whether the object is occluded. Figure~\ref{fig:vlm_example} shows an example prompt used for occlusion analysis. When occlusion is detected in an image, we calculate the occlusion ratio $\rho$, defined as:
\begin{equation}
\rho = \frac{A_{\text{other}}}{A_{\text{other}} + A_{\text{target}}}
\end{equation}
where $A_{\text{other}}$ represents the area of other objects and $A_{\text{target}}$ represents the area of the target object. When the occlusion ratio $\rho < 0.4$, we apply amodal completion model to process the image. Otherwise, we discard the image and randomly select another image from the same region for evaluation. If all images in a region $R_i$ fail to meet the requirements, we proceed with only the qualifying images from other regions for subsequent processing.

\begin{figure}[h!]
    \centering
    \includegraphics[width=1.0\linewidth, trim={0 0 0 0}, clip]{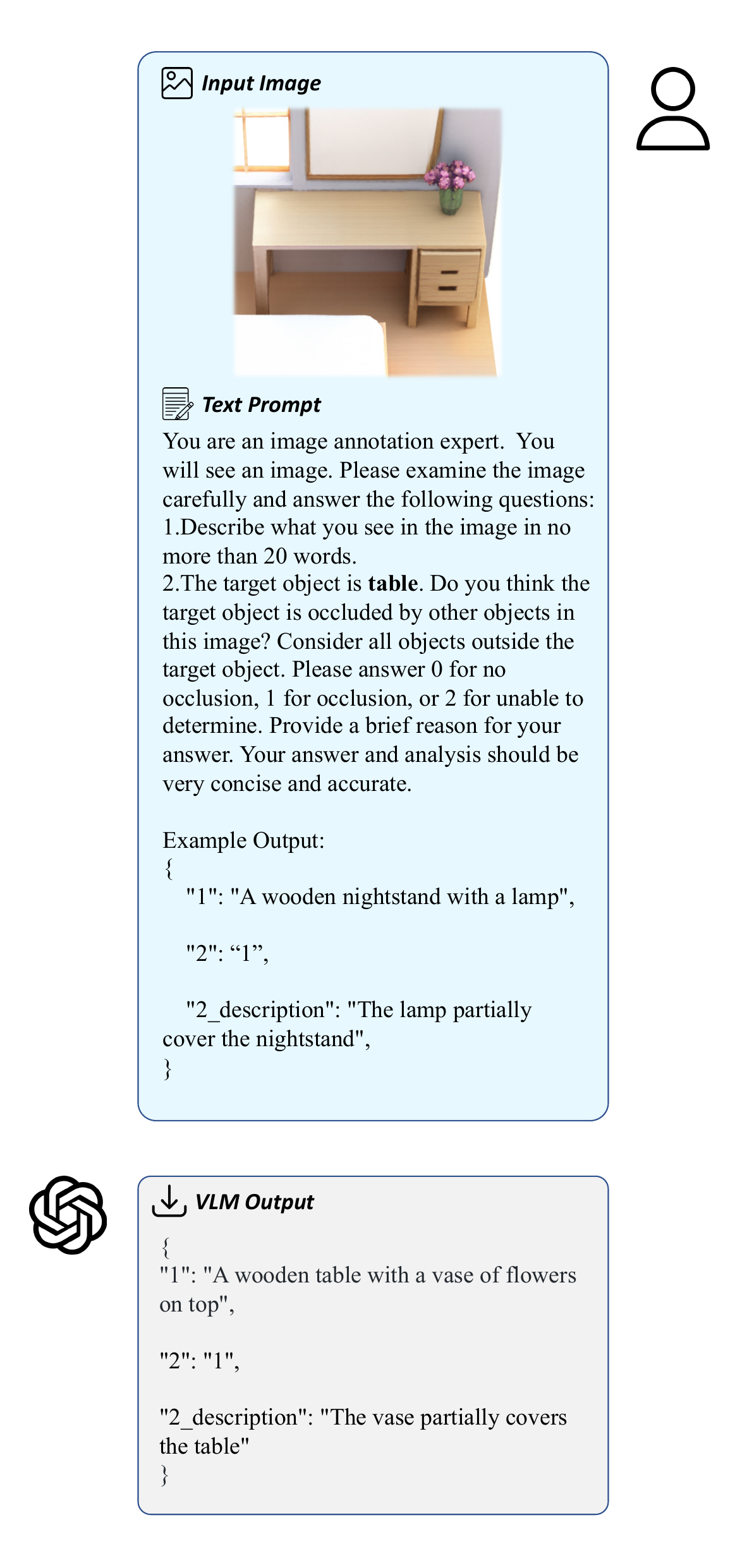}
    \caption{
    We provide an example of using VLM to determine whether the target object is occluded.
    }
    \label{fig:vlm_example}
\end{figure}

\begin{figure*}[h!]
    \centering
    \includegraphics[width=1.0\linewidth, trim={0 0 0 0}, clip]{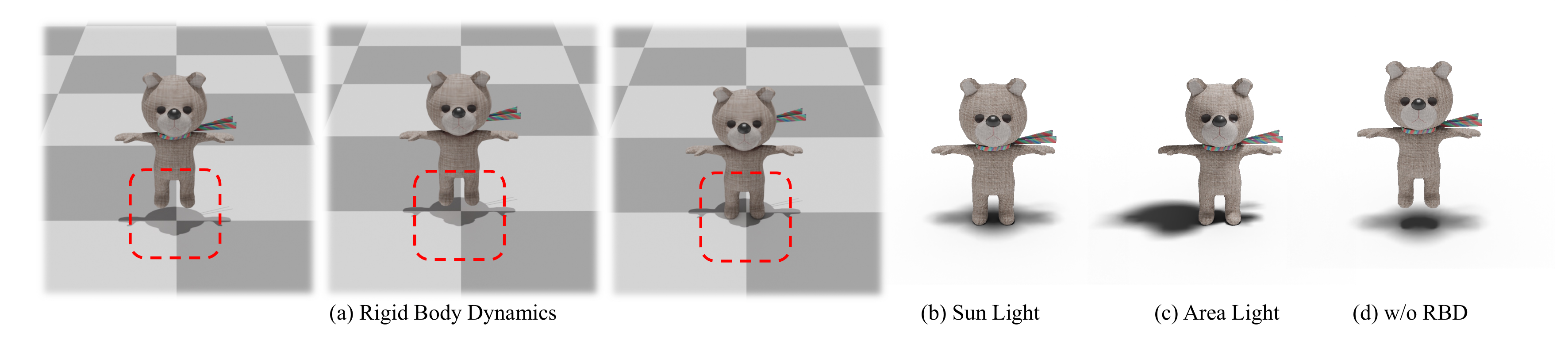}
    \caption{
    (a) Visualization of the Rigid Body Dynamics (RBD) process; (b) and (c) illustrate the shadow effects under different lighting setup; (d) demonstrates that incorrect shadow effects when object are not pre-processed with RBD.
    }
    \label{fig:render_example}
\end{figure*}

\section{Amodal Completion Implementation Details}
\label{sec:supp_amodal}

\subsection{Dataset Preparation}
Our large-scale amodal completion dataset is derived from two sources. Part of the data comes from previous work~\cite{ozguroglu2024pix2gestalt}, which is processed and annotated based on the SA-1B ~\cite{kirillov2023sam} dataset. Readers can refer to the original paper for more details. Additionally, to effectively handle visual effects (such as shadows) produced by occluding objects in real-world scenarios, we synthesized additional training data containing natural shadow effects.

\noindent\textbf{High-quality Shadow Data Synthesis Based on Objaverse.}
Objaverse~\cite{deitke2023objaverse}, as a large-scale diverse dataset containing over 800K 3D assets, provided us with rich 3D model resources. However, there are numerous low-quality models in this dataset, which often have overly simple geometric structures or missing textures. Therefore, we first strictly filtered Objaverse, ultimately obtaining approximately 181K high-quality 3D models for subsequent processing.

To generate realistic shadow effects, we used Blender\footnote{\url{https://www.blender.org}} Cycles rendering engine for high-quality rendering. In real-world scenarios, objects are typically placed on the ground or other supporting surfaces rather than floating in space, which is crucial for shadow formation. Simple coordinate normalization cannot solve the problem of object-ground contact, so we adopted a physics simulation approach: first normalizing the 3D models to the $[-1,1]^3$ spatial range, then adding Rigid Body Constraints to the object, and simulating the natural process of objects falling to the ground under gravity. In Figure~\ref{fig:render_example}, we present an example. We set the simulation time to 200 seconds and used the object's posture in the final stable state for rendering to ensure natural contact between the object and the ground.

\begin{algorithm}[ht]
\caption{Randomized Lighting Setup}
\label{alg:lighting_setup}
\begin{algorithmic}[1]
    \ENSURE Scene with randomized lighting configuration
    \vspace{1mm}
    \STATE Sample $p \sim \mathcal{U}(0, 1)$
    \IF{$p > 0.2$}
        \STATE \textbf{/* Area Light Setup */}
        \STATE Sample radius $R \sim \mathcal{U}(4.0, 6.0)$
        \STATE Sample energy $E \sim \mathcal{U}(800, 1200)$
        \STATE Sample size $S \sim \mathcal{U}(0.8, 1.2)$
        \STATE Sample elevation angle $\theta \sim \mathcal{U}(40°, 89.9°)$
        \STATE Sample azimuth angle $\phi \sim \mathcal{U}(0°, 360°)$
        \STATE Add area light $\mathcal{L}(R, E, S, \theta, \phi)$
    \ELSE
        \STATE \textbf{/* Sun Light Setup */}
        \STATE Sample sun angle $\alpha \sim \mathcal{U}(0.1, 0.5)$
        \STATE Add primary sun light with angle $\alpha$ and energy $5.0$
        \FOR{$i \in \{1, 2, 3\}$}
            \STATE Add sun light with angle $\alpha$ and energy $3.0$
            \STATE Rotate light by $\{90°, 180°, 270°\}[i-1]$ around x-axis
        \ENDFOR
    \ENDIF
\end{algorithmic}
\end{algorithm}

\begin{algorithm}[ht]
    \caption{Occlusion-to-Visibility Transition Generation}
    \label{alg:occlusion_transition}
    \begin{algorithmic}[1]
    \REQUIRE $M_v$: visible mask, $M_w$: whole mask, $I_o$: target RGB image, $N$: interpolation frames
    \ENSURE $\mathbf{X}$: video data
    \STATE $M_d \leftarrow |M_v - M_w|$ \COMMENT{Difference mask}
    \STATE $I_v \leftarrow I_o \odot M_w$ \COMMENT{Visible RGB}
    \STATE Initialize $\mathbf{X} \in \mathbb{R}^{(N+1) \times c \times h \times w}$ \COMMENT{Output tensor}
    \STATE $\mathbf{X}_0 \leftarrow \text{normalize}(\text{resize}(I_o, w \times h))$ \COMMENT{First frame}
    \FOR{$i \in \{0, 1, ..., N-1\}$}
        \STATE $\alpha_i = 1 - \frac{i}{N-1}$ \COMMENT{Blending coefficient}
        \STATE $M_i \leftarrow M_d \cdot \alpha_i$ \COMMENT{Weighted mask}
        \STATE $I_i \leftarrow \text{blend}(I_v, I_w, M_i)$ \COMMENT{Blended image}
        \STATE $\mathbf{X}_{i+1} \leftarrow \text{normalize}(\text{resize}(I_i, w \times h))$
    \ENDFOR
    \RETURN $\mathbf{X}$
    \end{algorithmic}
\end{algorithm}

\noindent\textbf{Lighting Setup.}
To simulate diverse real-world lighting conditions, we primarily used two types of light sources in Blender during the rendering process: Sun Light and Area Light. Sun Light simulates parallel light rays produced by distant light sources, with controllable lighting effects through parameters such as Strength and Angle; Area Light simulates light emitted from a plane, achieving different lighting effects by adjusting parameters such as Size and Power. As outlined in Algorithm\ref{alg:lighting_setup}, for each rendering, we randomly selected one of these light source types and randomly set relevant parameters to enhance data diversity.

After rendering, we adopted a method similar to previous method~\cite{ozguroglu2024pix2gestalt}, overlaying 3D object images with shadows onto complete scene images from SA-1B to form image pairs containing occlusion relationships. Through the above processing, we ultimately constructed a large-scale amodal completion dataset containing approximately 1.32 million data pairs, of which about 468K data pairs include natural shadow effects, providing rich training resources for the model to learn to process complex occlusion scenarios. As shown in Figure~\ref{fig:render_example}, the shadow effects under different lighting are illustrated.

\subsection{Network Architecture}
In Section 4, we proposed a amodal completion network based on Stable Video Diffusion~\cite{blattmann2023stablevideo}  (SVD) . This section elaborates on its network architecture design.

Stable Video Diffusionis an image-conditioned video generation network. Trained on massive video datasets, given an input image $I$, it can generate a temporally consistent video from $I$. The SVD architecture consists of three core components: a video encoder $\mathcal{E}$, a decoder $\mathcal{D}$, and a UNet diffusion model $\Phi$.

To fully leverage the prior knowledge embedded in the pre-trained model, we optimize the network structure following the principle of minimal modification. Specifically, the input image $I$ is first compressed through the encoder $\mathcal{E}$ to obtain an image embedding representation. Simultaneously, the object observation mask $M$ undergoes downsampling to maintain consistent resolution with the image embedding. In the original SVD architecture, the image embedding is concatenated with latent noise; in our improved version, we incorporate the mask $M$ into the concatenation process, providing additional information to better guide the video generation process. The image $I$ is also processed through CLIP to extract image features, which are injected via cross-attention. 

\subsection{Training Details}
In the model training process, we initialized our model with the pre-trained weights from SVD~\cite{blattmann2023stablevideo}. To accommodate the additional mask information, we expanded the input channels of thn illustration of the use diffusion model $\Phi$ from 8 to 9. In our experiments, video data was processed at a resolution of $512 \times 512$ with 8 frames per sequence. Training was conducted using fp16 precision with a learning rate of $1 \times 10^{-5}$ and incorporated a 500-step warmup phase. All parameters of $\Phi$ were fine-tuned during this process. We used a batch size of 8 and trained on 8 Nvidia H20 GPUs, with the entire training process taking approximately 3 days.

\section{Spatial Aligned Generation}
\label{supp_spatial_align}
Here we provide an algorithm~\ref{alg:spatial_aligned} for spatial aligned generation.

\begin{algorithm}[t!]
    \caption{Spatial Aligned Generation}
    \label{alg:spatial_aligned}
    \begin{algorithmic}[1]
    \REQUIRE Observation views $I$ with camera parameters $[r_i, p_i]$, timestep range $[t_{min}, t_{max}]$
    \ENSURE Generated 3D structure $\widehat{S}$
    \vspace{1mm}
    
    \STATE \textbf{/* Initial Structure Generation */}
    \STATE Generate initial geometric structure $S_{\text{init}} \leftarrow F_{\text{LRM}}($I$, [r_i, p_i])$
    \STATE Obtain explicit voxel representation $V$ through voxelization of $S_{\text{init}}$
    \STATE Compress $V$ into latent feature $S_{\text{implicit}} \leftarrow \mathcal{E}_S(V)$
    
    \STATE \textbf{/* Shape Prior Injection */}
    \STATE Select intermediate timestep $t \in [t_{min}, t_{max}]$ \COMMENT{Typically $t \in [0.2, 0.4]$}
    \STATE Initialize $x_t \leftarrow S_{\text{implicit}} + \text{noise}(t)$ \COMMENT{Add noise corresponding to timestep $t$}
    \STATE Set number of discretization steps $N$
    
    \STATE \textbf{/* Rectified Flow Sampling Process */}
    \FOR{$i = $INT$(N\times t-1)$ \textbf{down to} $0$}
        \STATE Compute current time $t_i$, step size $h$
        \STATE Predict velocity $v_\theta(x_{t_i}, t_i)$ using trained model $\boldsymbol{\mathcal{G}}_{\mathrm{S}}$
        \STATE Update sample $x_{i-1} = x_{i} - h \cdot v_\theta(x_{i}, t_i)$
    \ENDFOR
    \STATE Set generated structure $\widehat{S} \leftarrow x_0$
    \STATE \textbf{return} $\widehat{S}$
    \end{algorithmic}
\end{algorithm}

\section{Background Structure Generation}
\label{supp_background}
For each scene, we regenerate the background by modifying the original scene prompt to begin with 'A background of...', then integrate the resulting background structure back into the resulting scene.

\vspace{-0.5em}
\section{User Study}
\label{sup:userstudy}
Our user study evaluates the generated 3D scenes from two dimensions: image-text matching degree and overall scene quality. The matching degree assesses how well the generated 3D scene aligns with the input text description, while the overall quality comprehensively considers factors such as scene quality, reasonableness of object layout, and the quality of individual object models. As shown in Figure~\ref{fig:user_study}, we designed an intuitive user study interface where participants can quantitatively score the generated results. The scoring standard ranges from 1 to 3 points, with higher scores indicating better quality. Through this approach, we can qualitatively analyze the practical effectiveness of the generated results.
\vspace{-1.0em}
\begin{figure}[h!]
    \centering
    \includegraphics[width=1.0\linewidth, trim={0 0 0 0}, clip]{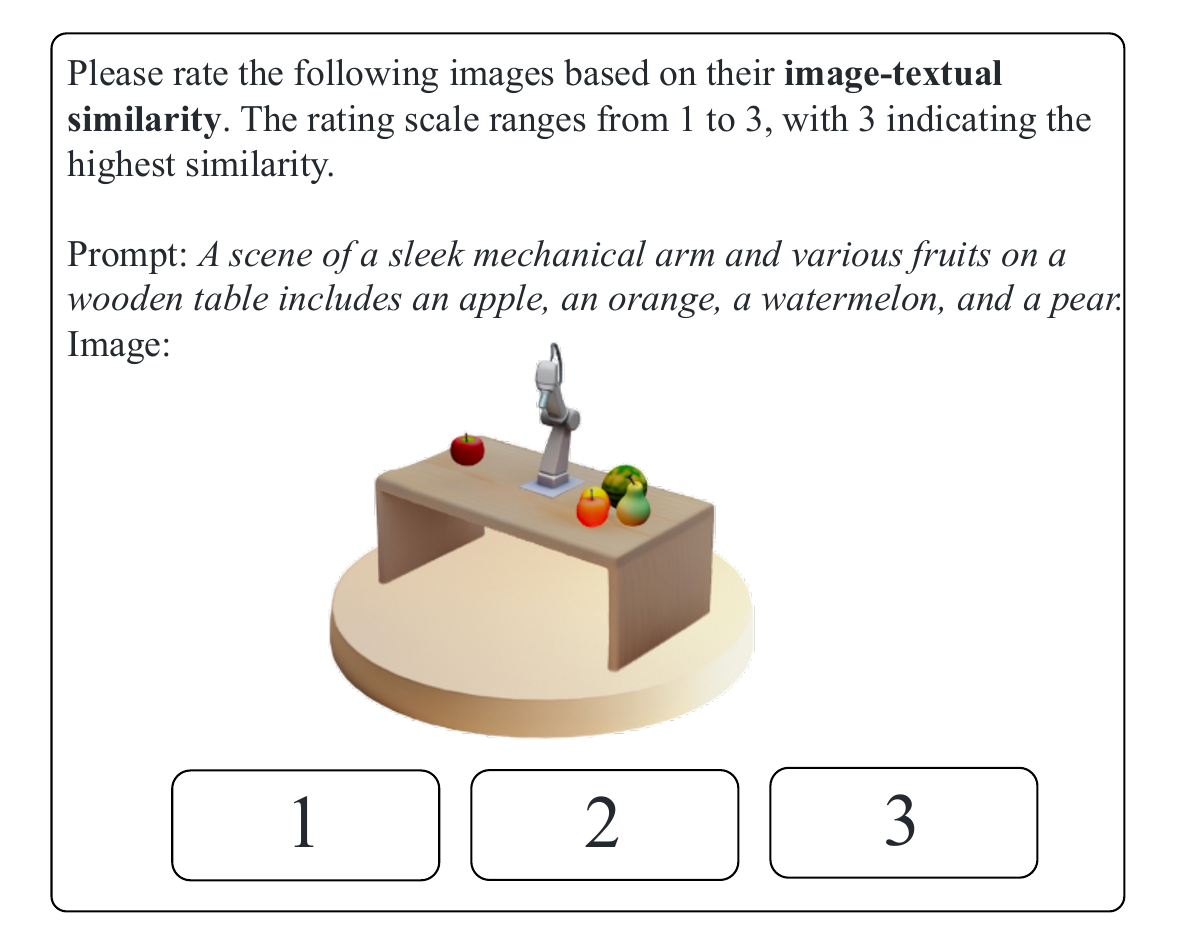}
    \caption{
    An illustration of the user study interface.
    }
    \vspace{-1.0em}
    \label{fig:user_study}
\end{figure}
\vspace{-1.0em}

\section{Runtime Evalution}
\label{sec:runtime_evaluation}
As shown in Figure 2 of the main paper, given a text prompt, HiScene first utilizes FLUX~\cite{flux2024} to generate candidate images within 4 seconds, followed by TRELLIS to complete scene initialization in 5 seconds. Next, HiScene renders multi-view images and performs 2D semantic segmentation in 5 seconds. In the 3D Gaussian semantic segmentation stage, we optimize 5000 steps, taking approximately 2 minutes. After obtaining the initial 3D GS representation of objects, the system spends less than 1 second rendering object candidate images and completes occlusion analysis in 30 seconds. For each candidate image with occlusion, amodal completion requires 6 seconds.
As illustrated in Figure 5 of the main paper, during the Spatial Alignment stage for each object, we first utilize Sparse view LRM to obtain initial geometric structures in 2 seconds, followed by voxelization in less than 1 second, and generate a low-resolution feature grid using $E_S$ in 1 second. Finally, the system takes 3 seconds for structure generation and structure latents generation.

Overall, HiScene processes a complete scene in approximately 12 minutes. In contrast, methods relying on SDS Loss optimization such as GALA3D and DreamScene require significantly more time. GALA3D needs 2 hours to generate a scene, while DreamScene requires more than 1 hour. This comparison clearly demonstrates HiScene's significant advantage in efficiently generating interactive 3D scenes.

\begin{figure*}[h!]
    \centering
    \includegraphics[width=0.92\linewidth, trim={0 0 0 0}, clip]{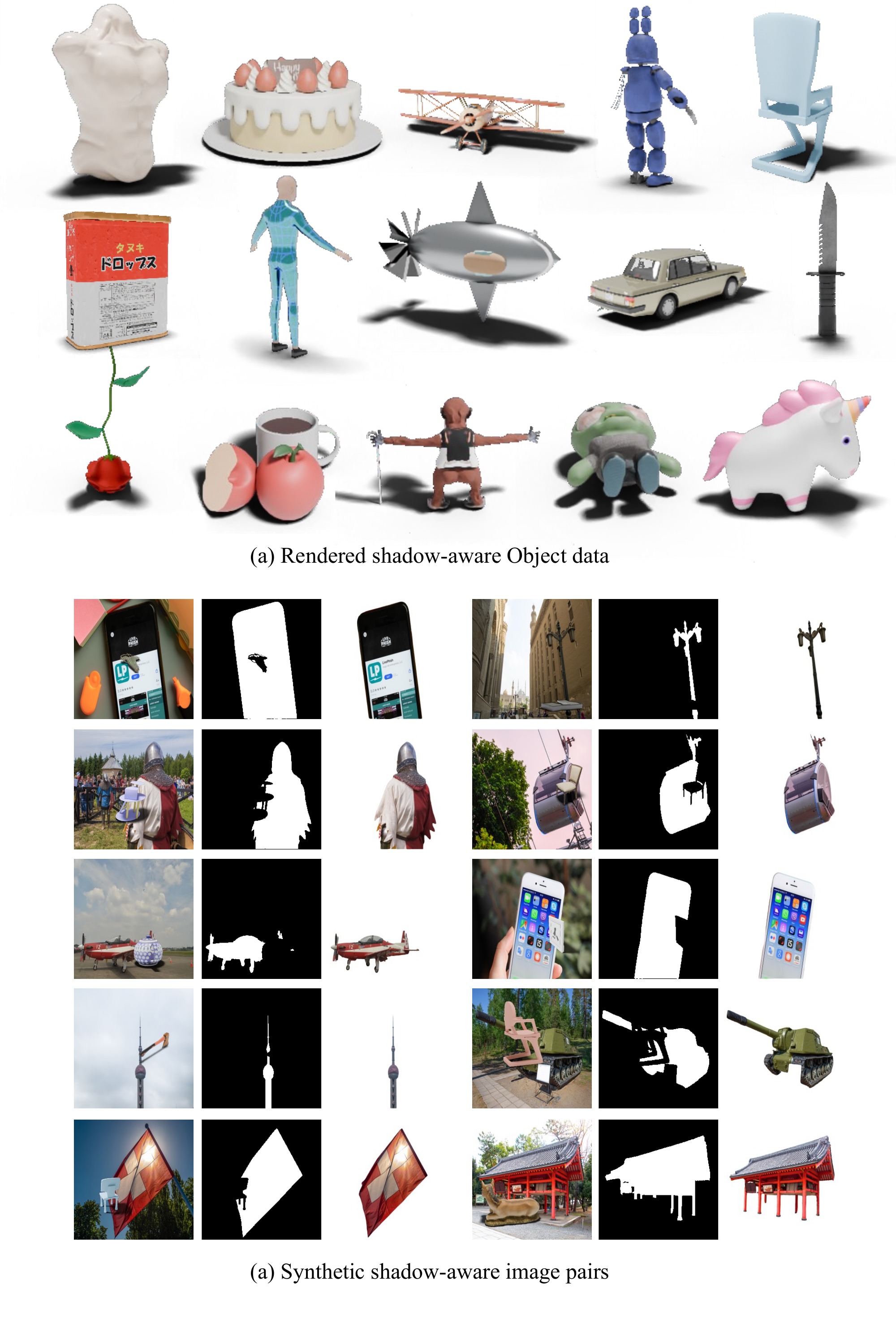}
    \caption{
    We show examples of our synthetic dataset.
    }
    \vspace{-4.0em}
    \label{fig:supp_example_shadowdata}
\end{figure*}

\begin{figure*}[h!]
    \centering
    \includegraphics[width=0.92\linewidth, trim={0 0 0 0}, clip]{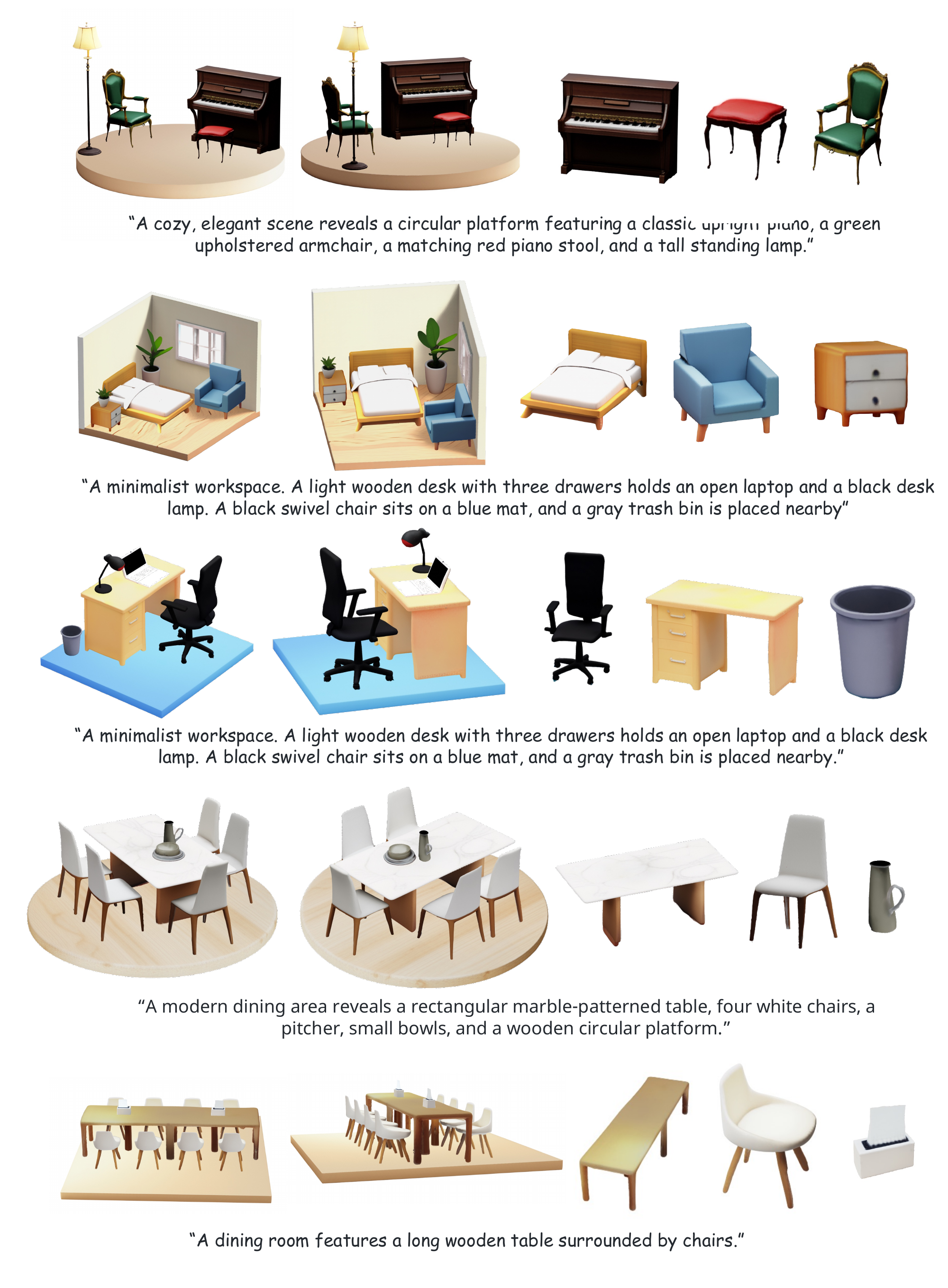}
    \caption{
    More examples of generated scenes. All prompts have a fixed prefix \textit{"Isometric view of "}.
    }
    \vspace{-4.0em}
    \label{fig:supp_more_example}
\end{figure*}

\begin{figure*}[h!]
    \centering
    \includegraphics[width=1.0\linewidth, trim={0 0 0 0}, clip]{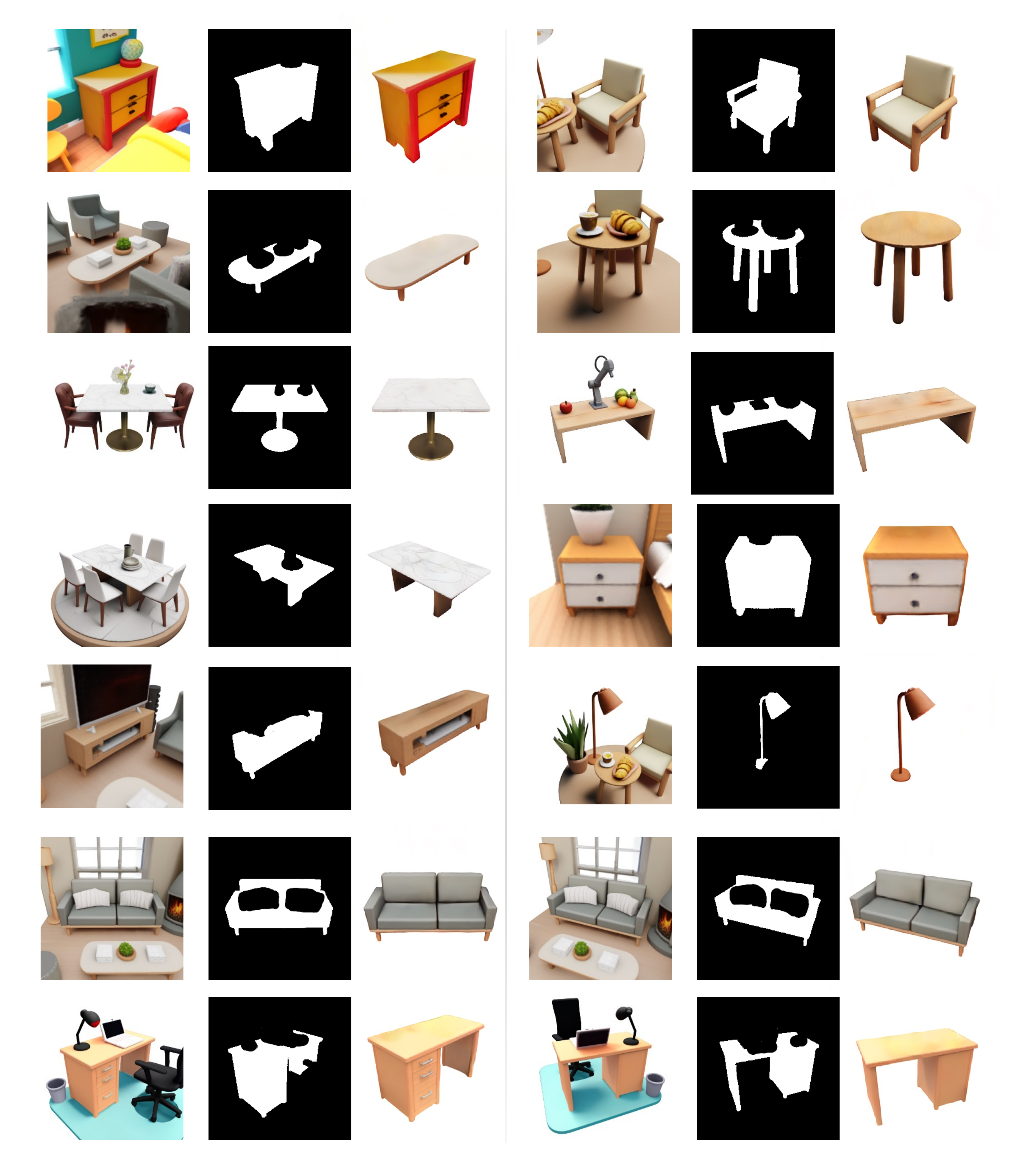}
    \caption{
    More examples of amodal completion. 
    }
    \vspace{-4.0em} 
    \label{fig:supp_example_amodal}
\end{figure*} \fi

\end{document}